\documentclass[10pt]{article}
\usepackage[a4paper,margin=1in]{geometry}
\usepackage[T1]{fontenc}
\usepackage[utf8]{inputenc}
\usepackage{lmodern}
\usepackage{microtype}
\usepackage{amsmath,amssymb}
\usepackage{braket}
\usepackage{graphicx}
\graphicspath{{./}{figures/}{regge_plots/}{ff_plots/}}
\usepackage{booktabs}
\usepackage{longtable}
\usepackage{multirow}
\usepackage{array}
\usepackage{subcaption}
\usepackage{cite}
\usepackage[hidelinks]{hyperref}
\usepackage{float}
\usepackage{placeins}
\usepackage{xcolor}
\usepackage{setspace}
\usepackage{ifthen}
\newcommand{\tablefont}{\fontsize{6.2}{7.0}\selectfont}

\title{Structure, Spectroscopy, and Decay Phenomenology of Charm--Strange Tetraquarks}
\author{Chetan Lodha\\Department of Physics, Sardar Vallabhbhai National Institute of Technology,\\Surat, Gujarat-395007, India\\\texttt{iamchetanlodha@gmail.com}
\and Ajay Kumar Rai\\Department of Physics, Sardar Vallabhbhai National Institute of Technology,\\Surat, Gujarat-395007, India\\\texttt{raiajayk@gmail.com}}
\date{}

\hypersetup{
  pdftitle={Structure, Spectroscopy, and Decay Phenomenology of Charm-Strange Tetraquarks},
  pdfauthor={Chetan Lodha and Ajay Kumar Rai},
  pdfsubject={Pauli-consistent spectroscopy, decays, and Regge trajectories of csbar{s}bar{s} tetraquarks},
  pdfkeywords={tetraquark, Pauli principle, charm-strange spectroscopy, diquark-antidiquark, Regge trajectories, PDG candidates}
}
\newcolumntype{L}[1]{>{\raggedright\arraybackslash}p{#1}}
\newcolumntype{C}[1]{>{\centering\arraybackslash}p{#1}}

\begin{document}
\maketitle

\begin{abstract}
We investigate the structure, spectroscopy, and decay phenomenology of charm--strange $cs\bar{s}\bar{s}$ tetraquarks within a compact diquark--antidiquark description. The analysis first constructs the allowed color--spin configurations of the identical-antistrange subsystem, relates them to local tetraquark currents and the Lorentz structures used in QCD sum-rule projections, and then evaluates the spectrum with a Cornell-type potential in parallel semi-relativistic (SR) and non-relativistic (NR) schemes. The same interaction is benchmarked against the conventional $D_s$ spectrum before being applied to the diquark and tetraquark sectors. The calculation covers $D_s$ states through $4S$, $4P$, $4D$, and $4F$ and tetraquark states through $3S$, $3P$, and $3D$, while a reduced set is displayed to emphasize the physically relevant excitation patterns. The compact tetraquark spectrum exhibits a persistent separation between the lighter antitriplet and heavier sextet color sectors, whereas the absolute masses and threshold locations show appreciable SR--NR dependence. Orbital and radial Regge trajectories are used as internal spectral diagnostics, and the lowest states are organized relative to the relevant $D_s^{(*)}\eta^{(\prime)}$ and $D_s^{(*)}\phi$ thresholds. The decay analysis proceeds from $D_s$ weak and hidden-strange benchmark channels, treated with form factors and factorization, to tetraquark rearrangement decays described by a fall-apart mechanism with explicit color--spin recoupling.
\end{abstract}


\section{Introduction}

Quantum chromodynamics (QCD) permits color-singlet configurations beyond conventional $q\bar q$ mesons and $qqq$ baryons~\cite{Gell-Mann:1964ewy,Zweig:1964jf,ParticleDataGroup:2026rpp}. The growing spectrum of hidden- and open-flavor structures has therefore motivated compact tetraquark, molecular, coupled-channel, hybrid, and threshold-based interpretations~\cite{Jaffe:1976ig,Jaffe:1976ih,Belle:2003nnu,BESIII:2013ris,LHCb:2014zfx,LHCb:2020bwg,LHCb:2020bls,LHCb:2020pxc,Guo:2017jvc,Berwein:2024ztx}. Open-flavor systems are especially informative because their conserved quantum numbers can exclude an ordinary quarkonium assignment and expose the roles of color correlations, Fermi statistics, and nearby two-hadron thresholds.

The charm--strange sector provides a sensitive testing ground. The low masses of $D_{s0}^{*}(2317)$ and $D_{s1}(2460)$ relative to early quark-model expectations have been discussed in terms of compact multiquark components, $DK$ or $D^{*}K$ molecules, and coupled-channel dynamics~\cite{BaBar:2003oey,CLEO:2003ggt,Belle:2003guh,Godfrey:1985xj,Faessler:2007us,Moir:2016srx,Guo:2017jvc}. More recently, LHCb has established the open-charm strange structures $T_{\bar c\bar s0}^{*}(2870)^0$, $T_{\bar c\bar s1}^{*}(2900)^0$, and the isovector $T_{c\bar s0}^{*}(2900)$ pair in reconstructed charm-meson channels~\cite{LHCb:2020bls,LHCb:2020pxc,LHCb:2024xyx,LHCb:2024vfz,LHCb:2023Tcbs2900}. Their flavor assignments differ from $cs\bar s\bar s$, but they demonstrate that open-flavor exotics can be isolated experimentally. The broad charged isoscalar resonance $D_{sJ}(3040)^+$ is included in the candidate audit of Sec.~\ref{subsec:tq-candidate-audit}~\cite{BaBar:2009rro,ParticleDataGroup:2026rpp}.

A controlled tetraquark analysis also requires a reliable conventional $c\bar s$ benchmark. Relativized and nonrelativistic potential models provide reference $D_s$ spectra and Regge systematics~\cite{Godfrey:1985xj,DiPierro:2001uu,Ebert:2009ua,Devlani:2011zz,Kher:2017mky,Kher:2021ifh}, while chiral, unitarized, molecular, and lattice-QCD approaches emphasize threshold effects in the positive-parity sector~\cite{Bardeen:2003kt,Kolomeitsev:2003ac,vanBeveren:2003kd,Faessler:2007us,Mohler:2013rwa,Lang:2014yfa,Moir:2016srx,Guo:2017jvc}. Compact open-charm tetraquarks have in parallel been studied with constituent potentials, diquark--antidiquark models, QCD sum rules, coupled-channel methods, and Born--Oppenheimer constructions~\cite{Jaffe:1976ig,Ebert:2010af,Shifman:1978bx,Shifman:1978by,Zhang:2006ix,Yang:2023evp,Wang:2023eng,Berwein:2024ztx}. These approaches differ dynamically, but all require a consistent treatment of color--spin symmetry, thresholds, and decay channels.

Related journal-published applications by the present author and collaborators span all-charm, all-light, light--strange and all-strange tetraquarks, hadronic molecules, open-charm and open-beauty bound states, and the $B_c$ meson~\cite{LodhaJCM:2023allcharm,LodhaRai:2024strangeonium,LodhaRai:2024alllight,LodhaRai:2025lightstrange,YadavLodhaRai:2025molecule,LodhaParmarRai:2025charm,LodhaParmarRai:2026beauty,PatelLodha:2026Bc}. They are cited here as methodological context across neighboring flavor sectors; only results directly relevant to the present $cs\bar s\bar s$ system are used in the comparisons below.

The $cs\bar s\bar s$ configuration is particularly clean because its leading rearrangement topology,
\[
cs\bar s\bar s\rightarrow(c\bar s)(s\bar s),
\]
selects $D_s^{(*)}\eta^{(\prime)}$ and $D_s^{(*)}\phi$ final states. The two identical antistrange quarks simultaneously impose a nontrivial restriction on the antidiquark color and spin. Section~\ref{subsec:pauli-basis} enforces this exchange symmetry before the spectrum, currents, decays, and Regge trajectories are constructed.

We first determine the conventional $D_s$ spectrum in parallel semi-relativistic (SR) and non-relativistic (NR) treatments and then apply the same interaction to the Pauli-filtered compact bases. The calculation itself is deliberately more extensive than the spectrum displayed in the main text: the $D_s$ calculation is performed through $4S$, $4P$, $4D$, and $4F$, while the tetraquark calculation is performed through $3S$, $3P$, and $3D$. To keep the physics visible, the main tables retain only $D_s$ levels through $4S$, $3P$, $2D$, and $1F$, and tetraquark levels through $3S$, $2P$, and $1D$. The omitted higher levels are retained as consistency and trajectory inputs rather than promoted to additional phenomenological predictions. The remaining outputs comprise $n=1$ spin-compatible thresholds, Regge fits, recoupling factors, reduced rearrangement coefficients, and one-daughter cascade weights. Because off-diagonal color--spin mixing is not evaluated, the tetraquark masses remain diagonal compact-basis values.

The results have different levels of robustness. The Pauli-allowed basis, the absence of the internal-$S$-wave sextet $2^+$ basis, and exact color--spin recoupling zeros follow from symmetry and angular-momentum algebra. By contrast, the numerical masses, SR--NR differences, threshold excesses, and Regge slopes depend on the chosen potential and kinematic prescription. The reduced coefficients additionally factor out an unknown channel-dependent strong coupling. Candidate identifications are therefore treated as the least model-independent layer of the analysis.

Section~\ref{sec:1} develops the compact basis and Hamiltonian. Section~\ref{sec:spectroscopy} presents the meson, diquark, and tetraquark spectra and threshold assignments; Sec.~\ref{sec:4} gives the Regge analysis; and Sec.~\ref{sec:decay} develops the meson benchmarks and tetraquark rearrangement framework. The phenomenological interpretation and candidate assessment are collected in Sec.~\ref{sec:results_discussion}.

\section{Theoretical framework}
\label{sec:theoretical-framework}
\label{sec:1}

A common Cornell-type central interaction, supplemented by perturbative spin-dependent terms, is applied to the $D_s$, diquark, and Pauli-filtered diquark--antidiquark sectors in parallel SR and NR treatments~\cite{Eichten:1978Cornell,Godfrey:1985xj,Lucha:1995zv,Ebert:2009ua}.

\subsection{Pauli constraint and construction of the compact basis}
\label{subsec:pauli-basis}

We denote the charm--strange diquark by $d=[cs]$ and the antidiquark formed from the two identical antistrange quarks by $\bar d=[\bar s\bar s]$, with cluster spins $S_d$ and $S_{\bar d}$. Color neutrality permits the two conjugate cluster couplings
\[
[cs]_{\bar{\mathbf 3}_c}[\bar s\bar s]_{\mathbf 3_c}
\qquad\text{or}\qquad
[cs]_{\mathbf 6_c}[\bar s\bar s]_{\bar{\mathbf 6}_c}.
\]
The Pauli condition acts on the identical antiquarks inside $\bar d$ and must be enforced before that antidiquark is coupled to the $[cs]$ cluster.

Under interchange $3\leftrightarrow4$ of the two $\bar s$ antiquarks, the complete internal wave function must be antisymmetric:
\begin{equation}
P_{34}^{\rm space}P_{34}^{\rm flavor}P_{34}^{\rm color}P_{34}^{\rm spin}=-1.
\label{eq:pauli-total}
\end{equation}
For the lowest antidiquark configuration, the internal orbital angular momentum is $\ell_{\bar d}=0$, so the spatial factor is symmetric, $P_{34}^{\rm space}=+1$. The identical-flavor state $\bar s\bar s$ is also symmetric, $P_{34}^{\rm flavor}=+1$. Equation~\eqref{eq:pauli-total} therefore becomes
\begin{equation}
P_{34}^{\rm color}P_{34}^{\rm spin}=-1,
\label{eq:color-spin-antisymmetry}
\end{equation}
and hence requires the color and spin factors to have opposite exchange symmetries.

The relevant color and spin decompositions for two antiquarks are standard consequences of color SU(3), spin coupling, and Fermi statistics in compact multiquark bases~\cite{Jaffe:1976ig,Jaffe:1976ih,Ebert:2010af}:
\begin{equation}
\bar{\mathbf 3}_c\otimes\bar{\mathbf 3}_c
 =\mathbf 3_c^{\rm A}\oplus\bar{\mathbf 6}_c^{\rm S},
\qquad
\frac{1}{2}\otimes\frac{1}{2}
 =1^{\rm S}\oplus0^{\rm A}.
\label{eq:color-spin-decomposition}
\end{equation}
The antisymmetric color triplet must consequently be combined with the symmetric spin triplet, whereas the symmetric color-$\bar{\mathbf 6}_c$ representation must be combined with the antisymmetric spin singlet:
\begin{equation}
\begin{aligned}
(\bar s\bar s)_{\mathbf 3_c^{\rm A}}&:\qquad S_{\bar d}=1,\\
(\bar s\bar s)_{\bar{\mathbf 6}_c^{\rm S}}&:\qquad S_{\bar d}=0.
\end{aligned}
\label{eq:pauli-allowed-antidiquarks}
\end{equation}
The complementary internal-$S$-wave assignments, $(\bar s\bar s)_{\mathbf 3_c}$ with $S_{\bar d}=0$ and $(\bar s\bar s)_{\bar{\mathbf 6}_c}$ with $S_{\bar d}=1$, violate the Pauli condition and are excluded.

No corresponding exchange restriction applies to the nonidentical charm and strange quarks in $d=[cs]$. Both $S_d=0$ and $S_d=1$ are therefore allowed in the color-$\bar{\mathbf 3}_c$ and color-$\mathbf 6_c$ diquark representations. Coupling each diquark to the conjugate antidiquark gives the allowed color--spin sectors
\begin{equation}
\begin{aligned}
\bar{\mathbf 3}_c\otimes\mathbf 3_c:&\qquad
S_{\bar d}=1,\quad S_d=0,1,\\
\mathbf 6_c\otimes\bar{\mathbf 6}_c:&\qquad
S_{\bar d}=0,\quad S_d=0,1.
\end{aligned}
\label{eq:allowed-color-spin-sectors}
\end{equation}
The cluster spins combine as $S_T=|S_d-S_{\bar d}|,\ldots,S_d+S_{\bar d}$. For relative diquark--antidiquark orbital angular momentum $L=0$, the parity is $P=(-1)^L=+$ and $J=S_T$. The $\bar{\mathbf 3}_c\otimes\mathbf 3_c$ sector therefore contains a scalar--axial configuration, $(S_d,S_{\bar d})=(0,1)$, with a single $S_T=1$ and $J^P=1^+$ basis, together with an axial--axial configuration, $(1,1)$, that produces $S_T=0,1,2$ and hence the $J^P=0^+,1^+,2^+$ multiplet. In the $\mathbf 6_c\otimes\bar{\mathbf 6}_c$ sector the antidiquark is necessarily scalar: $(0,0)$ gives $J^P=0^+$, whereas $(1,0)$ gives $J^P=1^+$.

These four cluster-spin families generate six independent $L=0$ basis states: two with $J^P=0^+$, three with $J^P=1^+$, and one with $J^P=2^+$. Thus no Pauli-allowed compact $S$-wave sextet--antisextet $2^+$ state exists, and the entire sextet axial-antidiquark family is absent from the spectrum, decay analysis, and Regge construction.

The orbital notation used below refers exclusively to relative motion between the diquark and antidiquark. The $P$- and $D$-wave tetraquarks have $L=1$ and $2$, respectively, but the two antiquarks remain internally in $\ell_{\bar d}=0$. Their spatial exchange parity therefore remains symmetric, so the same Pauli-allowed spin--color assignments apply to every radial and relative-orbital trajectory considered here. For these states,
\begin{equation}
P=(-1)^L,
\qquad
J=|L-S_T|,\,|L-S_T|+1,\ldots,L+S_T.
\label{eq:orbital-coupling-pauli-basis}
\end{equation}
An internal orbital excitation of the antidiquark would alter its spatial exchange symmetry and require a separate Pauli analysis; such configurations are outside the scope of the present cluster model.

The resulting color--spin configurations are basis vectors, not fully diagonalized physical resonances, as is generally the case before color--spin mixing is diagonalized in compact tetraquark treatments~\cite{Jaffe:1976ih,Ebert:2010af,Zhang:2006ix}. At $L=0$, off-diagonal color--spin interactions can mix the two $0^+$ bases and the three $1^+$ bases, while the $2^+$ basis is unique within the present compact space. A complete diagonalization would therefore require a $2\times2$ mass matrix in the $0^+$ sector and a $3\times3$ matrix in the $1^+$ sector. The numerical importance is potentially largest in the antitriplet $1^+$ sector: the two diagonal antitriplet $1^+$ levels differ by only $56.5$~MeV in SR and $91.9$~MeV in NR. For two basis states with diagonal masses $M_1,M_2$ and an off-diagonal matrix element $V_{12}$, the eigenvalues would be
\begin{equation}
M_{\pm}=\frac{M_1+M_2}{2}\pm\sqrt{\left(\frac{M_1-M_2}{2}\right)^2+|V_{12}|^2}.
\label{eq:mixing-sensitivity}
\end{equation}
Since $V_{12}$ and the corresponding three-state matrix elements are not evaluated here, no mixed eigenmass uncertainty can be assigned. The quoted masses are therefore diagonal compact-basis values rather than physical pole positions; conclusions involving an individual $0^+$ or $1^+$ mass must be read with this limitation in mind.

For compactness, the notation used below is fixed here in prose. We write $d=[cs]$ and $\bar d=[\bar s\bar s]$ for the diquark and antidiquark, with cluster spins $S_d$ and $S_{\bar d}$ coupled to $S_T$ before the relative orbital motion is added. The quantum numbers $L$, $n$, and $J^P$ denote the relative diquark--antidiquark orbital angular momentum, radial level, and total spin--parity, respectively. ``SR'' denotes the semi-relativistic kinetic expansion through $\mathcal O(p^6)$ and ``NR'' the non-relativistic kinetic term. The signed excess $\Delta_X=M_T^X-M_{\rm th}$ is only a kinematic distance from threshold in scheme $X$, whereas $K_\alpha$ in $\Gamma_\alpha=K_\alpha|\gamma_\alpha|^2$ contains phase space and color--spin--flavor recoupling but not the channel-dependent strong coupling.

\subsection{Pauli-allowed local currents and sum-rule correspondence}
\label{subsec:local-currents}

The same Pauli projection can be expressed directly at the operator level. This is useful for connecting the compact potential-model basis to two- and three-point QCD sum-rule constructions, where the interpolating current fixes the quantum numbers and Lorentz structures of the correlator~\cite{Shifman:1978bx,Shifman:1978by,Yang:2023evp,Chen:2020fierz}.  We define the antisymmetric and symmetric color tensors
\begin{equation}
{\cal A}^{bc,de}=\epsilon^{abc}\epsilon^{ade},
\qquad
{\cal S}^{bc,de}=\delta^{bd}\delta^{ce}+\delta^{be}\delta^{cd},
\label{eq:color-tensors-currents}
\end{equation}
where an overall normalization of ${\cal S}^{bc,de}$ may be absorbed into the pole residue.  The four Pauli-allowed antitriplet currents are
\begin{align}
J_{0,\bar3}^{AA}={}&{\cal A}^{bc,de}
[c_b^TC\gamma_\alpha s_c][\bar s_d\gamma^\alpha C\bar s_e^T],
\label{eq:current-aa0}\\
J_{1\mu,\bar3}^{SA}={}&{\cal A}^{bc,de}
[c_b^TC\gamma_5s_c][\bar s_d\gamma_\mu C\bar s_e^T],
\label{eq:current-sa1}\\
J_{1\mu\nu,\bar3}^{AA}={}&{\cal A}^{bc,de}\Big(
[c_b^TC\gamma_\mu s_c][\bar s_d\gamma_\nu C\bar s_e^T]
-[c_b^TC\gamma_\nu s_c][\bar s_d\gamma_\mu C\bar s_e^T]\Big),
\label{eq:current-aa1}\\
J_{2\mu\nu,\bar3}^{AA}={}&{\cal A}^{bc,de}\Big(
[c_b^TC\gamma_\mu s_c][\bar s_d\gamma_\nu C\bar s_e^T]
+[c_b^TC\gamma_\nu s_c][\bar s_d\gamma_\mu C\bar s_e^T]
-\tfrac12 g_{\mu\nu}[c_b^TC\gamma_\alpha s_c]
[\bar s_d\gamma^\alpha C\bar s_e^T]\Big).
\label{eq:current-aa2}
\end{align}
They represent, respectively, the $(1,1);0$, $(0,1);1$, $(1,1);1$, and $(1,1);2$ cluster-spin bases.  The two Pauli-allowed sextet currents are
\begin{align}
J_{0,6}^{SS}={}&{\cal S}^{bc,de}
[c_b^TC\gamma_5s_c][\bar s_d\gamma_5C\bar s_e^T],
\label{eq:current-ss0}\\
J_{1\mu,6}^{AS}={}&{\cal S}^{bc,de}
[c_b^TC\gamma_\mu s_c][\bar s_d\gamma_5C\bar s_e^T],
\label{eq:current-as1}
\end{align}
which correspond to $(0,0);0$ and $(1,0);1$.  No additional local internally $S$-wave currents survive the exchange condition in Eq.~\eqref{eq:pauli-allowed-antidiquarks}. Equations~\eqref{eq:current-aa0}--\eqref{eq:current-as1} establish the operator correspondence for the compact basis; derivative currents would be required for QCD-sum-rule studies of the relative $P$- and $D$-wave levels. The numerical spectroscopy below is obtained from the potential model.

\subsection{QCD sum-rule projection structures}
\label{subsec:qcdsr-projections}

The Pauli-filtered operators also determine the Lorentz structures that must be isolated in current-specific two-point and three-point sum rules~\cite{Shifman:1978bx,Shifman:1978by,Yang:2023evp}. For a current $J_{\mathcal I}$, the two-point correlation function is
\begin{equation}
\Pi_{\mathcal I}(p)=i\int d^4x\,e^{ip\cdot x}
\langle0|T\{J_{\mathcal I}(x)J_{\mathcal I}^{\dagger}(0)\}|0\rangle.
\label{eq:qcdsr-two-point}
\end{equation}
For scalar currents the invariant amplitude is obtained directly. A vector current is decomposed as
\begin{equation}
\Pi_{\mu\nu}(p)=\left(-g_{\mu\nu}+\frac{p_\mu p_\nu}{p^2}\right)\Pi_1(p^2)
+\frac{p_\mu p_\nu}{p^2}\Pi_L(p^2),
\label{eq:qcdsr-vector-projector}
\end{equation}
so that the transverse invariant $\Pi_1$ isolates the spin-one pole. For the symmetric tensor current, the spin-two contribution is selected by
\begin{equation}
{\cal P}_{\mu\nu,\alpha\beta}=\frac12
(\widetilde g_{\mu\alpha}\widetilde g_{\nu\beta}
+\widetilde g_{\mu\beta}\widetilde g_{\nu\alpha})
-\frac13\widetilde g_{\mu\nu}\widetilde g_{\alpha\beta},
\qquad
\widetilde g_{\mu\nu}=-g_{\mu\nu}+\frac{p_\mu p_\nu}{p^2}.
\label{eq:qcdsr-spin2-projector}
\end{equation}
The antisymmetric axial--axial current requires a rank-four projection rather than the ordinary vector projector. In the positive-parity convention,
\begin{equation}
\langle0|J_{1\mu\nu,\bar3}^{AA}|1^+(p)\rangle
=\lambda_+\epsilon_{\mu\nu\rho\sigma}\epsilon^\rho p^\sigma,
\label{eq:qcdsr-aa1-pole}
\end{equation}
and the corresponding epsilon--epsilon tensor structure must be selected to avoid opposite-parity contamination. For strong decays, the three-point correlator introduced in Sec.~\ref{subsec:three-point-requirements} must be projected consistently for the scalar, vector, antisymmetric-tensor, or symmetric-tensor current before the double Borel transformation and physical-pole extrapolation~\cite{Shifman:1978bx,Shifman:1978by,Yang:2023evp}. These projectors provide the operator-level bridge from the compact basis to current-specific masses, residues, and strong couplings. A literature survey for this revision did not identify a completed OPE for the exact $cs\bar s\bar s$ flavor and all six currents in Eqs.~\eqref{eq:current-aa0}--\eqref{eq:current-as1}; existing sum-rule calculations cited below therefore serve as analogous-current comparisons rather than current-specific inputs.

\subsection{Hamiltonian and spin-dependent interactions}
The masses and binding energies are obtained from a modified time-independent radial Schr\"odinger equation in the center-of-mass frame, following the standard potential-model treatment of heavy-quark bound states~\cite{Eichten:1978Cornell,Lucha:1995zv,Godfrey:1985xj}. The SR and NR Hamiltonians are
\begin{equation}
    H_{SR} = \sum_{i=1}^{2}\sqrt{M_i^2+p_i^2}+V^{(0)}(r)+V_{SD}(r),
\end{equation}
\begin{equation}
    H_{NR} = \sum_{i=1}^{2}\left(\frac{p_i^2}{2M_i}+M_i\right)+V^{(0)}(r)+V_{SD}(r),
\end{equation}
where $M_i$ and $p_i$ are the constituent masses and relative momenta, while $V^{(0)}(r)$ and $V_{SD}(r)$ denote the central and spin-dependent interactions.

In the SR treatment, the kinetic contribution is expanded through $\mathcal{O}(p^6)$,
\begin{equation}
    K.E.=\sum_{i=1}^{2}\left[\frac{p^{2}}{2M_i}-\frac{p^{4}}{8M_i^{3}}+\frac{p^{6}}{16M_i^{5}}\right],
\end{equation}
which provides a controlled higher-order correction for systems containing both charm and strange constituents. The central interaction is taken to be the Cornell Coulomb-plus-linear potential~\cite{Eichten:1978Cornell,Lucha:1995zv,Godfrey:1985xj}
\begin{equation}
    V_{C+L}^{(0)}(r)=\frac{k_s\alpha_s}{r}+br+V_0,
\end{equation}
where $k_s$ is the color factor, $\alpha_s$ a fixed effective coupling for the fitted charm--strange parameter set, $b$ the string tension, and $V_0$ a phenomenological constant. The complete central potential is
\begin{equation}
    V^{(0)}(r)=V_{C+L}^{(0)}(r)+V^{1}(r)\left(\frac{1}{m_1}+\frac{1}{m_2}\right),
\end{equation}
with the leading mass-dependent correction
\begin{equation}
    V^{1}(r)=-\frac{C_F C_A}{4}\frac{\alpha_s^2}{r^2},
\end{equation}
where $C_F=4/3$ and $C_A=3$~\cite{Koma:2006si,Koma:2012bc}.

The numerical inputs used in the $c\bar s$ calibration and propagated to the $[cs][\bar s\bar s]$ calculation are stated here to keep the parameter set visible without a separate table. In the SR scheme we use $b=0.1300$~GeV$^2$, $V_0=-0.210$~GeV, $\sigma=0.9500$~GeV, $\alpha_s=0.6000$, $m_c=1.475$~GeV, and $m_s=0.550$~GeV. The corresponding NR inputs are $b=0.1300$~GeV$^2$, $V_0=-0.215$~GeV, $\sigma=1.1500$~GeV, $\alpha_s=0.6000$, $m_c=1.480$~GeV, and $m_s=0.525$~GeV. The coupling $\alpha_s$ is a fitted effective constant rather than an explicitly scale-running quantity, while $\sigma$ is the Gaussian smearing scale of the contact hyperfine term in Eq.~\eqref{eqss} and phenomenologically regularizes the short-distance spin--spin interaction.

The spin-dependent interaction is added perturbatively and separated into tensor, spin--orbit, and spin--spin contributions, using the usual Breit--Fermi organization employed in quark-potential spectroscopy~\cite{Godfrey:1985xj,Koma:2006si,Koma:2012bc},
\begin{equation}
    V_{SD}(r)=V_T(r)+V_{LS}(r)+V_{SS}(r),
\end{equation}
whose Breit--Fermi forms are
\begin{subequations}
\begin{equation}
\left(-\frac{k_s\alpha_s}{4}\frac{12\pi}{M_{\mathcal{D}}M_{\bar{\mathcal{D}}}}\frac{1}{r^3}\right)
\left(-\frac{1}{3}(S_1\cdot S_2)+\frac{(S_1\cdot r)(S_2\cdot r)}{r^2}\right),
\label{eqvt}
\end{equation}
\begin{equation}
\left(-\frac{3\pi k_s\alpha_s}{2M_{\mathcal{D}}M_{\bar{\mathcal{D}}}}\frac{1}{r^3}-\frac{b}{2M_{\mathcal{D}}M_{\bar{\mathcal{D}}}}\frac{1}{r}\right)(L\cdot S),
\label{eqls}
\end{equation}
\begin{equation}
\left(-\frac{k_s\alpha_s}{3}\frac{8\pi}{M_{\mathcal{D}}M_{\bar{\mathcal{D}}}}\left(\frac{\sigma}{\sqrt{\pi}}\right)^3 e^{-\sigma^2r^2}\right)(S_1\cdot S_2).
\label{eqss}
\end{equation}
\end{subequations}
These contributions generate the fine and hyperfine splittings among states with the same orbital structure but different spin couplings. The same interaction scheme has been used in our earlier open-charm and multiquark calculations~\cite{LodhaRai:2024strangeonium,LodhaRai:2024alllight,LodhaRai:2025lightstrange,LodhaParmarRai:2025charm,LodhaParmarRai:2026beauty,PatelLodha:2026Bc} and is applied here consistently to the charm--strange meson, diquark, and tetraquark sectors.

\subsection{Uncertainty convention and model limitations}
\label{subsec:uncertainty-limits}
The uncertainties attached to the tabulated masses are the propagated input/model variations used in the underlying two-stage diquark and tetraquark calculation. They should not be interpreted as a complete statistical error budget. In particular, they do not include correlations among the fitted Cornell parameters, terms beyond $\mathcal O(p^6)$ in the SR expansion, higher-order treatment of $V_{SD}$, off-diagonal color--spin mixing, or hadronic-loop and coupled-channel self-energies; the importance of the latter effects is well established in heavy-light spectroscopy near open thresholds~\cite{Guo:2017jvc,Mohler:2013rwa,Lang:2014yfa,Moir:2016srx}. The SR--NR separation is therefore quoted throughout as a separate diagnostic of kinematic-scheme dependence rather than folded into the stated $\pm$ uncertainty.

The spin-dependent terms are treated in first-order perturbation theory. A posteriori, the axial--axial $S$-wave multiplet corresponds to an effective spin--spin scale of about $81.6$~MeV (SR) and $109.8$~MeV (NR), inferred from the $0^+,1^+,2^+$ splittings. This is useful as a measure of fine-structure size, but the present data set does not retain a state-by-state value of $\langle V_{SD}\rangle/\langle V^{(0)}\rangle$; we therefore do not claim a quantitative perturbative-convergence test. The missing ratio is included among the model limitations rather than reconstructed from the final masses.

\section{Spectroscopy}
\label{sec:spectroscopy}

The spectroscopy calculation has three distinct purposes. First, the conventional $c\bar s$ spectrum calibrates the charm--strange mass scale and supplies the daughter masses that enter the tetraquark thresholds. Second, the diquark calculation provides Pauli-consistent correlated $[cs]$ and $[\bar s\bar s]$ building blocks. Third, those ingredients are combined into the compact $cs\bar s\bar s$ spectrum, whose phenomenological relevance is assessed relative to physical two-meson thresholds rather than by mass values alone. This ordering is important because the three levels of output do not have the same status: the low-lying $D_s$ spectrum is a calibration test, the diquark masses are model-dependent intermediate quantities, and the absolute tetraquark masses inherit both parameter and kinematic uncertainties. The discussion below therefore emphasizes the spectral patterns that survive these limitations instead of treating every tabulated number as an equally robust prediction.

\subsection{Meson spectrum: calibration of the charm--strange sector}

The $D_s$ system is the natural calibration sector because the same charm--strange interaction subsequently enters both the $[cs]$ diquark and the $D_s^{(*)}\eta^{(\prime)}$, $D_s^{(*)}\phi$ rearrangement thresholds. Conventional potential models and lattice calculations provide a well-developed reference for this spectrum~\cite{Godfrey:1985xj,Ebert:2009ua,Devlani:2011zz,Godfrey:2015dva,Ni:2021pce,Mohler:2013rwa,Lang:2014yfa}. For a color-singlet meson, $\bar{\mathbf 3}\otimes\mathbf 3=\mathbf 1\oplus\mathbf 8$ and the singlet color coefficient is $k_s=-4/3$. The mass is evaluated as
\begin{equation}
    M_{(c\bar{s})}=M_c+M_{\bar{s}}+E_{(c\bar{s})}+\braket{V^{0}(r)},
\end{equation}
with the kinetic prescription and spin-dependent terms specified in Sec.~\ref{sec:theoretical-framework}. The purpose of this step is not to refit each excited $D_s$ level independently, but to test whether one common interaction reproduces the ground-state scale and maintains a credible radial and orbital ordering.

The numerical solution was obtained through $4S$, $4P$, $4D$, and $4F$. Table~\ref{mass_meson}, however, is intentionally restricted to $4S$, $3P$, $2D$, and $1F$. Retaining all four $S$-wave radial levels gives the longest experimentally anchored radial sequence and is useful for testing the curvature of the radial spacing. Three $P$-wave levels are sufficient to expose the low-lying fine structure and its radial repetition; two $D$-wave levels establish the first higher-orbital recurrence; and the lowest $F$-wave multiplet completes the $n=1$ orbital progression through $L=3$. The still higher $P$-, $D$-, and $F$-wave excitations are increasingly remote from experimental constraints and add little to the conclusions drawn from the main spectrum.

The entries are quoted in the unmixed $LS$ basis. This convention is useful for keeping the spin-dependent Hamiltonian transparent, but it should not be confused with a complete physical-state analysis: heavy-light states with the same $J^P$, notably $n\,{}^{1}P_1$--$n\,{}^{3}P_1$, $n\,{}^{1}D_2$--$n\,{}^{3}D_2$, and $n\,{}^{1}F_3$--$n\,{}^{3}F_3$, can mix~\cite{Godfrey:1985xj,Ebert:2009ua,Godfrey:2015dva,Ni:2021pce}. Consequently, agreement with individual positive-parity masses is a diagnostic of the model rather than a one-to-one state-identification criterion.

\begingroup
\tablefont
\setlength{\tabcolsep}{0.35pt}
\renewcommand{\arraystretch}{0.92}
\setlength{\LTcapwidth}{\textwidth}
\begin{longtable}{@{}>{\centering\arraybackslash}p{16mm}>{\centering\arraybackslash}p{9mm}*{2}{>{\centering\arraybackslash}p{11mm}}*{8}{>{\centering\arraybackslash}p{13mm}}@{}}
\caption{Selected central $D_s$ meson masses in MeV. The numerical calculation extends through $4S$, $4P$, $4D$, and $4F$; the main-text subset is limited to $4S$, $3P$, $2D$, and $1F$ so that the table retains the calibration and ordering information without turning the spectroscopy section into an exhaustive level catalogue. Present SR and NR results are compared with Particle Data Group central values~\cite{ParticleDataGroup:2026rpp} and representative theoretical calculations.}
\label{mass_meson}\\
\toprule
\multirow{2}{*}{State} & \multirow{2}{*}{$J^P$} & \multicolumn{2}{c}{Present} & \multicolumn{8}{c}{Comparison} \\
\cmidrule(lr){3-4}\cmidrule(l){5-12}
 & & SR & NR & \cite{ParticleDataGroup:2026rpp} & \cite{Devlani:2011zz} & \cite{Kher:2017mky} & \cite{Patel:2021ftu} & \cite{Ni:2021pce} & \cite{Ebert:2009ua} & \cite{Godfrey:2015dva} & \cite{Jakhad:2025zos} \\
\midrule
\endfirsthead
\multicolumn{12}{c}{\tablename\ \thetable\ -- continued from previous page}\\
\toprule
\multirow{2}{*}{State} & \multirow{2}{*}{$J^P$} & \multicolumn{2}{c}{Present} & \multicolumn{8}{c}{Comparison} \\
\cmidrule(lr){3-4}\cmidrule(l){5-12}
 & & SR & NR & \cite{ParticleDataGroup:2026rpp} & \cite{Devlani:2011zz} & \cite{Kher:2017mky} & \cite{Patel:2021ftu} & \cite{Ni:2021pce} & \cite{Ebert:2009ua} & \cite{Godfrey:2015dva} & \cite{Jakhad:2025zos} \\
\midrule
\endhead
\midrule
\multicolumn{12}{r}{Continued on next page}\\
\endfoot
\bottomrule
\endlastfoot
\multicolumn{12}{@{}c@{}}{$\mathbf{L=0}$ states}\\
\midrule
\(1\,{}^{1}S_{0}\) & \(0^{-}\) & 1967.9 & 2003.5 & 1968.35 & 1970 & 1965 & 1966 & 1969 & 1969 & 1979 & 1969 \\
\(2\,{}^{1}S_{0}\) & \(0^{-}\) & 2560.1 & 2599.2 & $2591^{\dagger}$ & 2688 & 2680 & 2645 & 2649 & 2688 & 2673 & 2709 \\
\(3\,{}^{1}S_{0}\) & \(0^{-}\) & 2955.9 & 3004.2 & -- & 3158 & 3247 & 3118 & 3126 & 3219 & 3154 & 3139 \\
\(4\,{}^{1}S_{0}\) & \(0^{-}\) & 3298.5 & 3342.8 & -- & 3556 & 3764 & 3487 & -- & 3652 & 3547 & 3481 \\
\(1\,{}^{3}S_{1}\) & \(1^{-}\) & 2113.1 & 2128.7 & 2112.2 & 2117 & 2120 & 2119 & 2112 & 2111 & 2129 & 2087 \\
\(2\,{}^{3}S_{1}\) & \(1^{-}\) & 2637.1 & 2665.5 & 2714 & 2723 & 2719 & 2683 & 2737 & 2731 & 2732 & 2714 \\
\(3\,{}^{3}S_{1}\) & \(1^{-}\) & 3010.9 & 3054.5 & -- & 3180 & 3265 & 3135 & 3191 & 3242 & 3193 & 3139 \\
\(4\,{}^{3}S_{1}\) & \(1^{-}\) & 3336.1 & 3384.8 & -- & 3571 & 3775 & 3496 & -- & 3669 & 3575 & 3481 \\
\multicolumn{12}{@{}c@{}}{$\mathbf{L=1}$ states}\\
\midrule
\(1\,{}^{1}P_{1}\) & \(1^{+}\) & 2474.8 & 2495.4 & 2535.11 & 2530 & 2529 & 2534 & 2528 & 2536 & 2549 & 2524 \\
\(2\,{}^{1}P_{1}\) & \(1^{+}\) & 2866.8 & 2904.6 & -- & 3019 & 3081 & 3015 & 3002 & 3067 & 3018 & 3019 \\
\(3\,{}^{1}P_{1}\) & \(1^{+}\) & 3198.6 & 3247.2 & -- & -- & 3587 & 3401 & -- & 3618 & 3416 & 3384 \\
\(1\,{}^{3}P_{0}\) & \(0^{+}\) & 2382.7 & 2402.3 & 2317.8 & 2444 & 2438 & 2436 & 2409 & 2509 & 2484 & 2461 \\
\(2\,{}^{3}P_{0}\) & \(0^{+}\) & 2780.7 & 2817.3 & -- & 2947 & 3022 & 2958 & 2940 & 3054 & 3005 & 3003 \\
\(3\,{}^{3}P_{0}\) & \(0^{+}\) & 3114.1 & 3162.3 & -- & -- & 3541 & 3358 & -- & 3513 & 3412 & 3377 \\
\(1\,{}^{3}P_{1}\) & \(1^{+}\) & 2466.0 & 2485.6 & 2459.5 & 2540 & 2541 & 2485 & 2545 & 2574 & 2556 & 2535 \\
\(2\,{}^{3}P_{1}\) & \(1^{+}\) & 2858.7 & 2895.3 & -- & 3023 & 3092 & 2987 & 3026 & 3154 & 3038 & 3009 \\
\(3\,{}^{3}P_{1}\) & \(1^{+}\) & 3189.8 & 3237.9 & -- & -- & 3596 & 3379 & -- & 3519 & 3433 & 3373 \\
\(1\,{}^{3}P_{2}\) & \(2^{+}\) & 2512.0 & 2531.6 & 2569.1 & 2566 & 2569 & 2548 & 2575 & 2571 & 2592 & 2572 \\
\(2\,{}^{3}P_{2}\) & \(2^{+}\) & 2905.4 & 2942.0 & -- & 3048 & 3109 & 3022 & 3053 & 3142 & 3048 & 3015 \\
\(3\,{}^{3}P_{2}\) & \(2^{+}\) & 3237.2 & 3285.3 & -- & -- & 3609 & 3405 & -- & 3580 & 3439 & 3374 \\
\multicolumn{12}{@{}c@{}}{$\mathbf{L=2}$ states}\\
\midrule
\(1\,{}^{1}D_{2}\) & \(2^{-}\) & 2737.0 & 2765.0 & -- & 2816 & 2853 & 2844 & 2857 & 2931 & 2900 & 2883 \\
\(2\,{}^{1}D_{2}\) & \(2^{-}\) & 3077.6 & 3121.4 & -- & 3312 & 3368 & 3262 & 3267 & 3403 & 3298 & 3278 \\
\(1\,{}^{3}D_{1}\) & \(1^{-}\) & 2729.4 & 2757.3 & $2859^{\ddagger}$ & 2873 & 2882 & 2866 & 2843 & 2913 & 2899 & 2857 \\
\(2\,{}^{3}D_{1}\) & \(1^{-}\) & 3065.8 & 3109.5 & -- & 3292 & 3394 & 3280 & 3233 & 3383 & 3306 & 3268 \\
\(1\,{}^{3}D_{2}\) & \(2^{-}\) & 2740.7 & 2768.7 & -- & 2896 & 2872 & 2836 & 2911 & 2961 & 2926 & 2869 \\
\(2\,{}^{3}D_{2}\) & \(2^{-}\) & 3079.9 & 3123.6 & -- & 3248 & 3384 & 3258 & 3306 & 3456 & 3323 & 3259 \\
\(1\,{}^{3}D_{3}\) & \(3^{-}\) & 2738.5 & 2766.4 & 2860 & 2834 & 2860 & 2841 & 2882 & 2971 & 2917 & 2886 \\
\(2\,{}^{3}D_{3}\) & \(3^{-}\) & 3082.5 & 3126.2 & -- & 3263 & 3372 & 3258 & 3299 & 3469 & 3311 & 3264 \\
\multicolumn{12}{@{}c@{}}{$\mathbf{L=3}$ states}\\
\midrule
\(1\,{}^{1}F_{3}\) & \(3^{+}\) & 2949.1 & 2986.0 & -- & -- & -- & 3095 & 3123 & 3254 & 3186 & 3164 \\
\(1\,{}^{3}F_{2}\) & \(2^{+}\) & 2962.2 & 2999.0 & -- & -- & -- & 3122 & 3176 & 3230 & 3208 & 3148 \\
\(1\,{}^{3}F_{3}\) & \(3^{+}\) & 2954.9 & 2991.7 & -- & -- & -- & 3096 & 3205 & 3266 & 3218 & 3138 \\
\(1\,{}^{3}F_{4}\) & \(4^{+}\) & 2937.4 & 2974.3 & -- & -- & -- & 3080 & 3134 & 3300 & 3190 & 3149 \\\end{longtable}
\endgroup

The most important calibration result is the ground-state scale. The SR calculation reproduces the $1\,{}^{1}S_0$ and $1\,{}^{3}S_1$ masses to approximately the MeV level, whereas the NR values are systematically higher by several tens of MeV. This establishes that the common parameter set can reproduce the masses that dominate the physical tetraquark thresholds, while the SR--NR difference provides a direct measure of the kinematic sensitivity that should be expected when the same Hamiltonian is applied to more weakly constrained excitations.

The orbital spectrum gives a second, less stringent check. The calculated $P$-, $D$-, and $F$-wave levels follow the expected increase with orbital excitation, but their absolute positions show a larger spread relative to experiment and to other models~\cite{Godfrey:1985xj,Ebert:2009ua,Devlani:2011zz,Mohler:2013rwa,Lang:2014yfa}. In particular, discrepancies among the unmixed $P$-wave states should be read together with the omitted heavy-light mixing and possible coupled-channel effects. The physically useful conclusion is therefore not that each high excitation is predicted precisely, but that the interaction fixes the low-energy $c\bar s$ scale while preserving a conventional radial and orbital hierarchy.

The symbols $\dagger$ and $\ddagger$ in Table~\ref{mass_meson} mark the PDG-listed $D_{s0}(2590)^+$ and $D_{s1}^{*}(2860)^+$, shown only as possible $2\,{}^{1}S_0$ and $1\,{}^{3}D_1$ benchmarks~\cite{ParticleDataGroup:2026rpp,LHCb:2020Ds0,LHCb:2014Ds2860}. Their detailed phenomenological interpretation is deferred to Sec.~\ref{subsec:results_ds_meson}, where configuration mixing and the size of the SR--NR spread can be considered explicitly.

\subsection{Diquark spectrum: Pauli-consistent building blocks}

The diquark masses are not observable hadron masses; they are correlated two-body inputs used to construct the compact tetraquark basis. We employ the standard color decomposition $\mathbf 3\otimes\mathbf 3=\bar{\mathbf 3}\oplus\mathbf 6$ and $\bar{\mathbf 3}\otimes\bar{\mathbf 3}=\mathbf 3\oplus\bar{\mathbf 6}$, as in conventional diquark--antidiquark treatments~\cite{Jaffe:1976ig,Jaffe:1976ih,Ebert:2010af,Zhang:2006ix}. The antitriplet diquark channel carries $k_s=-2/3$, while the sextet carries $k_s=+1/3$. The corresponding masses are
\begin{subequations}
\begin{align}
M_{ss}&=2M_s+E_{ss}+\langle V^0\rangle,\\
M_{cs}&=M_c+M_s+E_{cs}+\langle V^0\rangle.
\end{align}
\end{subequations}
The identical $ss$ pair is subjected to the Pauli projection derived in Sec.~\ref{sec:theoretical-framework}; thus a dash in Table~\ref{diquark} denotes a forbidden internal-$S$-wave spin--color configuration, not a failed or omitted calculation. The uncertainties follow the input/model-variation prescription of Sec.~\ref{subsec:uncertainty-limits}.

\begin{table}[!htbp]
\centering
\caption{Pauli-filtered scalar and axial diquark masses in MeV; the corresponding antidiquark masses are identical.}
\label{diquark}
\tablefont
\begin{tabular}{@{}ccrrrr@{}}
\toprule
Flavor & $J^P$ & $\bar{\mathbf 3}$ SR & $\bar{\mathbf 3}$ NR & $\mathbf 6$ SR & $\mathbf 6$ NR\\
\midrule
$ss$ & $0^+$ & -- & -- & $1559.82\pm15.26$ & $1359.83\pm15.02$ \\
$ss$ & $1^+$ & $1340.22\pm16.14$ & $1192.61\pm15.92$ & -- & -- \\
$cs$ & $0^+$ & $2258.40\pm25.49$ & $2138.68\pm26.12$ & $2489.15\pm25.78$ & $2312.36\pm26.07$ \\
$cs$ & $1^+$ & $2285.61\pm26.15$ & $2158.01\pm26.42$ & $2483.86\pm28.79$ & $2309.83\pm29.22$
\\
\bottomrule
\end{tabular}
\end{table}

Table~\ref{diquark} makes the Pauli structure immediately visible. For the identical $ss$ pair, the color-$\bar{\mathbf 3}$ channel survives only with spin one, whereas the color-$\mathbf 6$ channel survives only with spin zero. The nonidentical $cs$ pair is not subject to that restriction, so both scalar and axial configurations occur in both color representations. Within the present Hamiltonian the sextet $cs$ correlations are heavier than the corresponding antitriplet ones, while the NR treatment lowers all of the diquark masses relative to SR. These differences feed directly into the compact tetraquark spectrum, but they should be interpreted as model ingredients rather than separately measurable states.

\subsection{Tetraquark spectrum: color hierarchy and combined excitation pattern}

After the Pauli projection, the compact color-singlet basis consists of $\bar{\mathbf 3}_c\otimes\mathbf 3_c$ and $\mathbf 6_c\otimes\bar{\mathbf 6}_c$ cluster couplings. In the cluster convention used here their intercluster color factors are $k_s=-4/3$ and $-10/3$, respectively~\cite{Jaffe:1976ih,Ebert:2010af,Zhang:2006ix}. Since the Coulomb term is written as $V_C=k_s\alpha_s/r$ in Eq.~\eqref{eqvt}, both negative coefficients represent attraction in this convention. A diagonal compact basis state is assigned the mass
\begin{equation}
M_{cs\bar{s}\bar{s}}=M_{cs}+M_{\bar{s}\bar{s}}+E_{cs\bar{s}\bar{s}}+\langle V^0\rangle,
\end{equation}
and is labelled by $\left|S_d,S_{\bar d};S_T,L\right\rangle_J$, with parity $P=(-1)^L$.

The calculation was carried out through $3S$, $3P$, and $3D$ for every Pauli-allowed family, but the main spectrum is deliberately restricted to $3S$, $2P$, and $1D$. All three $S$-wave radial levels are retained because the radial spacing is visibly nonuniform and all three points enter the radial Regge analysis. The $2P$ level is sufficient to establish the first radial recurrence of each $P$-wave multiplet, whereas the lowest $D$-wave multiplet is the only $D$-wave input required for the $n=1$ orbital progression and threshold comparison. The higher $3P$, $2D$, and $3D$ towers introduce no new Pauli-allowed families and do not alter the qualitative SR--NR or color-sector hierarchy. Table~\ref{tab:spd-wave-masses} therefore combines the displayed $S$-, $P$-, and $D$-wave states in a single spectrum rather than splitting the same physical message over several tables.

\begingroup
\tablefont
\setlength{\tabcolsep}{0.7pt}
\renewcommand{\arraystretch}{0.96}
\setlength{\LTcapwidth}{\textwidth}
\begin{longtable}{@{}L{28mm}C{10mm}C{18mm}C{18mm}@{\hspace{2mm}}L{28mm}C{10mm}C{18mm}C{18mm}@{}}
\caption{Combined Pauli-allowed diagonal compact-basis $cs\bar s\bar s$ masses in MeV. The calculation was carried out through $3S$, $3P$, and $3D$ for every allowed family, whereas the main-text display is restricted to $3S$, $2P$, and $1D$. This subset is sufficient to establish the radial pattern, the first $P$-wave radial recurrence, and the lowest orbital sequence used in the threshold and Regge analyses. The two color sectors are shown side by side.}
\label{tab:spd-wave-masses}\\
\toprule
\multicolumn{4}{c}{$\bar{\mathbf3}_c\otimes\mathbf3_c$} & \multicolumn{4}{c}{$\mathbf6_c\otimes\bar{\mathbf6}_c$}\\
\cmidrule(lr){1-4}\cmidrule(lr){5-8}
Basis and $J^P$ & State & $M_{\rm SR}$ & $M_{\rm NR}$ & Basis and $J^P$ & State & $M_{\rm SR}$ & $M_{\rm NR}$\\
\midrule
\endfirsthead
\multicolumn{8}{c}{\tablename\ \thetable\ -- continued from previous page}\\
\toprule
\multicolumn{4}{c}{$\bar{\mathbf3}_c\otimes\mathbf3_c$} & \multicolumn{4}{c}{$\mathbf6_c\otimes\bar{\mathbf6}_c$}\\
\cmidrule(lr){1-4}\cmidrule(lr){5-8}
Basis and $J^P$ & State & $M_{\rm SR}$ & $M_{\rm NR}$ & Basis and $J^P$ & State & $M_{\rm SR}$ & $M_{\rm NR}$\\
\midrule
\endhead
\midrule
\multicolumn{8}{r}{Continued on next page}\\
\endfoot
\bottomrule
\endlastfoot
$(0,1),\ 1^+$ & $1\,{}^3S_1$ & $3089.2\pm36.4$ & $2919.7\pm33.9$ & $(0,0),\ 0^+$ & $1\,{}^1S_0$ & $4399.2\pm57.3$ & $4052.4\pm52.6$ \\
$(0,1),\ 1^+$ & $2\,{}^3S_1$ & $3691.0\pm47.6$ & $3514.9\pm44.6$ & $(0,0),\ 0^+$ & $2\,{}^1S_0$ & $4543.2\pm61.8$ & $4205.7\pm55.6$ \\
$(0,1),\ 1^+$ & $3\,{}^3S_1$ & $4061.7\pm33.1$ & $3900.7\pm29.0$ & $(0,0),\ 0^+$ & $3\,{}^1S_0$ & $4677.7\pm46.4$ & $4349.6\pm42.9$ \\
$(1,1),\ 0^+$ & $1\,{}^1S_0$ & $2951.1\pm47.3$ & $2717.9\pm43.5$ & $(1,0),\ 1^+$ & $1\,{}^3S_1$ & $4393.9\pm39.7$ & $4049.8\pm35.9$ \\
$(1,1),\ 0^+$ & $2\,{}^1S_0$ & $3663.1\pm46.8$ & $3449.1\pm43.8$ & $(1,0),\ 1^+$ & $2\,{}^3S_1$ & $4538.0\pm44.0$ & $4203.2\pm40.6$ \\
$(1,1),\ 0^+$ & $3\,{}^1S_0$ & $4051.9\pm37.7$ & $3861.0\pm34.4$ & $(1,0),\ 1^+$ & $3\,{}^3S_1$ & $4672.5\pm58.2$ & $4347.2\pm53.6$ \\
$(1,1),\ 1^+$ & $1\,{}^3S_1$ & $3032.7\pm47.3$ & $2827.8\pm43.5$ & & & & \\
$(1,1),\ 1^+$ & $2\,{}^3S_1$ & $3689.7\pm46.8$ & $3491.0\pm43.8$ & & & & \\
$(1,1),\ 1^+$ & $3\,{}^3S_1$ & $4069.3\pm37.7$ & $3889.7\pm34.4$ & & & & \\
$(1,1),\ 2^+$ & $1\,{}^5S_2$ & $3195.8\pm47.3$ & $3047.4\pm43.5$ & & & & \\
$(1,1),\ 2^+$ & $2\,{}^5S_2$ & $3742.9\pm46.8$ & $3574.8\pm43.8$ & & & & \\
$(1,1),\ 2^+$ & $3\,{}^5S_2$ & $4104.1\pm37.7$ & $3947.1\pm34.4$ & & & & \\
\addlinespace[2pt]
$(0,1),\ 0^-$ & $1\,{}^3P_0$ & $3506.4\pm48.0$ & $3316.1\pm44.6$ & $(0,0),\ 1^-$ & $1\,{}^1P_1$ & $4429.5\pm46.7$ & $4086.4\pm42.7$ \\
$(0,1),\ 0^-$ & $2\,{}^3P_0$ & $3898.6\pm43.8$ & $3723.6\pm42.0$ & $(0,0),\ 1^-$ & $2\,{}^1P_1$ & $4572.4\pm49.2$ & $4238.9\pm45.2$ \\
$(0,1),\ 1^-$ & $1\,{}^3P_1$ & $3543.6\pm48.0$ & $3353.6\pm44.6$ & $(1,0),\ 0^-$ & $1\,{}^3P_0$ & $4435.0\pm47.9$ & $4096.9\pm43.6$ \\
$(0,1),\ 1^-$ & $2\,{}^3P_1$ & $3931.1\pm43.8$ & $3756.8\pm42.0$ & $(1,0),\ 0^-$ & $2\,{}^3P_0$ & $4579.1\pm55.2$ & $4250.9\pm49.9$ \\
$(0,1),\ 2^-$ & $1\,{}^3P_2$ & $3617.9\pm48.0$ & $3428.6\pm44.6$ & $(1,0),\ 1^-$ & $1\,{}^3P_1$ & $4429.6\pm47.9$ & $4090.4\pm43.6$ \\
$(0,1),\ 2^-$ & $2\,{}^3P_2$ & $3996.1\pm43.8$ & $3823.5\pm42.0$ & $(1,0),\ 1^-$ & $2\,{}^3P_1$ & $4573.1\pm55.2$ & $4243.7\pm49.9$ \\
$(1,1),\ 1^-$ & $1\,{}^1P_1$ & $3587.5\pm55.9$ & $3392.8\pm51.2$ & $(1,0),\ 2^-$ & $1\,{}^3P_2$ & $4418.9\pm47.9$ & $4077.4\pm43.6$ \\
$(1,1),\ 1^-$ & $2\,{}^1P_1$ & $3971.3\pm43.5$ & $3790.7\pm39.7$ & $(1,0),\ 2^-$ & $2\,{}^3P_2$ & $4561.2\pm55.2$ & $4229.3\pm49.9$ \\
$(1,1),\ 0^-$ & $1\,{}^3P_0$ & $3462.8\pm55.9$ & $3264.2\pm51.2$ & & & & \\
$(1,1),\ 0^-$ & $2\,{}^3P_0$ & $3864.4\pm43.5$ & $3679.7\pm39.7$ & & & & \\
$(1,1),\ 1^-$ & $1\,{}^3P_1$ & $3590.0\pm55.9$ & $3394.7\pm51.2$ & & & & \\
$(1,1),\ 1^-$ & $2\,{}^3P_1$ & $3973.3\pm43.5$ & $3792.9\pm39.7$ & & & & \\
$(1,1),\ 2^-$ & $1\,{}^3P_2$ & $3627.9\pm55.9$ & $3432.2\pm51.2$ & & & & \\
$(1,1),\ 2^-$ & $2\,{}^3P_2$ & $4007.3\pm43.5$ & $3827.2\pm39.7$ & & & & \\
$(1,1),\ 1^-$ & $1\,{}^5P_1$ & $3462.5\pm55.9$ & $3261.8\pm51.2$ & & & & \\
$(1,1),\ 1^-$ & $2\,{}^5P_1$ & $3864.9\pm43.5$ & $3679.8\pm39.7$ & & & & \\
$(1,1),\ 2^-$ & $1\,{}^5P_2$ & $3620.8\pm55.9$ & $3423.6\pm51.2$ & & & & \\
$(1,1),\ 2^-$ & $2\,{}^5P_2$ & $4001.1\pm43.5$ & $3820.9\pm39.7$ & & & & \\
$(1,1),\ 3^-$ & $1\,{}^5P_3$ & $3677.8\pm55.9$ & $3479.8\pm51.2$ & & & & \\
$(1,1),\ 3^-$ & $2\,{}^5P_3$ & $4052.1\pm43.5$ & $3872.4\pm39.7$ & & & & \\
\addlinespace[2pt]
$(0,1),\ 1^+$ & $1\,{}^3D_1$ & $3829.9\pm46.0$ & $3646.4\pm42.3$ & $(0,0),\ 2^+$ & $1\,{}^1D_2$ & $4478.3\pm56.3$ & $4140.6\pm51.6$ \\
$(0,1),\ 2^+$ & $1\,{}^3D_2$ & $3840.0\pm46.0$ & $3656.1\pm42.3$ & $(1,0),\ 1^+$ & $1\,{}^3D_1$ & $4484.9\pm59.9$ & $4152.1\pm54.6$ \\
$(0,1),\ 3^+$ & $1\,{}^3D_3$ & $3855.2\pm46.0$ & $3670.6\pm42.3$ & $(1,0),\ 2^+$ & $1\,{}^3D_2$ & $4477.0\pm59.9$ & $4142.8\pm54.6$ \\
$(1,1),\ 2^+$ & $1\,{}^1D_2$ & $3868.5\pm42.4$ & $3677.8\pm39.9$ & $(1,0),\ 3^+$ & $1\,{}^3D_3$ & $4465.2\pm59.9$ & $4128.8\pm54.6$ \\
$(1,1),\ 1^+$ & $1\,{}^3D_1$ & $3849.2\pm42.4$ & $3658.4\pm39.9$ & & & & \\
$(1,1),\ 2^+$ & $1\,{}^3D_2$ & $3867.6\pm42.4$ & $3676.8\pm39.9$ & & & & \\
$(1,1),\ 3^+$ & $1\,{}^3D_3$ & $3877.6\pm42.4$ & $3685.9\pm39.9$ & & & & \\
$(1,1),\ 0^+$ & $1\,{}^5D_0$ & $3831.5\pm42.4$ & $3640.2\pm39.9$ & & & & \\
$(1,1),\ 1^+$ & $1\,{}^5D_1$ & $3843.0\pm42.4$ & $3651.9\pm39.9$ & & & & \\
$(1,1),\ 2^+$ & $1\,{}^5D_2$ & $3861.3\pm42.4$ & $3670.2\pm39.9$ & & & & \\
$(1,1),\ 3^+$ & $1\,{}^5D_3$ & $3879.7\pm42.4$ & $3688.2\pm39.9$ & & & & \\
$(1,1),\ 4^+$ & $1\,{}^5D_4$ & $3887.8\pm42.4$ & $3694.6\pm39.9$ & & & &

\end{longtable}
\endgroup

Three features of Table~\ref{tab:spd-wave-masses} are more informative than the individual high-lying masses. First, the surviving $\bar{\mathbf 3}_c\otimes\mathbf 3_c$ states are systematically much lighter than the $\mathbf 6_c\otimes\bar{\mathbf 6}_c$ states. This separation is substantially larger than the fine splittings within a given multiplet and persists in both SR and NR, so it is the clearest scheme-stable hierarchy produced by the present Hamiltonian. Second, within the antitriplet axial--axial $1S$ family the ordering $0^+<1^+<2^+$ is unchanged by the kinematic prescription even though the absolute masses shift appreciably. Third, increasing $n$ or $L$ raises the spectrum in the expected direction while compressing the higher radial spacings; this is why the complete displayed $3S$ sequence carries more information than another set of very high orbital excitations.

The $P$- and $D$-wave results should therefore be read mainly as tests of continuity of the compact spectrum. The two $P$-wave levels show that the first radial excitation preserves the same broad color hierarchy, while the $1D$ multiplets complete the lowest orbital sequence used in the Regge analysis. Their fine splittings depend on the spin-dependent interaction and on the SR/NR prescription and are correspondingly less robust than the gross antitriplet--sextet separation, consistent with the spread among other constituent and QCD-sum-rule tetraquark calculations~\cite{Ebert:2010af,Zhang:2006ix,Yang:2023evp}.

\subsection{Threshold organization of the lowest states}

A compact mass becomes phenomenologically meaningful only after it is compared with the physical hadronic channels to which the state can rearrange. For each $n=1$ $S$-, $P$-, and $D$-wave basis state we therefore use the lowest ground-meson channel with a nonzero leading spin-recoupling coefficient in the same total-spin class. This threshold-first organization is standard for near-threshold multiquark spectroscopy, where a mass prediction by itself does not determine whether a state is kinematically open or likely to remain narrow~\cite{Guo:2017jvc,Chen:2020fierz}. We define
\begin{equation}
M_{\rm th}=m_{H_1}+m_{H_2},\qquad
\Delta_X=M_T^X-M_{\rm th},\quad X=\mathrm{SR},\mathrm{NR},
\label{eq:spd-threshold-delta}
\end{equation}
where $\Delta_X>0$ denotes an open threshold and $\Delta_X<0$ a subthreshold compact-basis mass. The threshold energy itself is independent of the final relative partial wave; the latter controls the centrifugal suppression above threshold. For the leading fall-apart assignment we take $\ell_{\rm rel}=L$ and include the strange-flavor component of the physical $\eta$.

Using PDG 2026 meson masses~\cite{ParticleDataGroup:2026rpp}, the relevant physical onsets are $D_s^+\eta$ (2516.2~MeV), $D_s^{*+}\eta$ (2660.1~MeV), $D_s^+\eta'$ (2926.1~MeV), $D_s^+\phi$ (2987.8~MeV), $D_s^{*+}\eta'$ (3070.0~MeV), and $D_s^{*+}\phi$ (3131.7~MeV). The leading spin-class assignments are $D_s\eta$ for $S_T=0$, $D_s^*\eta$ for $S_T=1$, and $D_s^*\phi$ for $S_T=2$. Table~\ref{tab:threshold-deltas-n1} gives the corresponding $n=1$ threshold excesses for both kinematic prescriptions.

\begingroup
\tablefont
\setlength{\tabcolsep}{0.4pt}
\renewcommand{\arraystretch}{0.92}
\setlength{\LTcapwidth}{\textwidth}
\begin{longtable}{@{}C{15mm}C{7mm}C{12.5mm}C{13.2mm}C{9.3mm}C{9.3mm}C{9.3mm}@{\hspace{1mm}}C{15mm}C{7mm}C{12.5mm}C{13.2mm}C{9.3mm}C{9.3mm}C{9.3mm}@{}}
\caption{Lowest spin-compatible physical two-meson thresholds and signed excess energies for the Pauli-allowed $n=1$ states, arranged side by side for the $\bar{\mathbf3}_c\otimes\mathbf3_c$ and $\mathbf6_c\otimes\bar{\mathbf6}_c$ sectors. Positive $\Delta_X$ denotes a mass above threshold, whereas a negative value denotes a subthreshold compact-basis state.}
\label{tab:threshold-deltas-n1}\\
\toprule
\multicolumn{7}{c}{$\bar{\mathbf3}_c\otimes\mathbf3_c$} & \multicolumn{7}{c}{$\mathbf6_c\otimes\bar{\mathbf6}_c$}\\
\cmidrule(lr){1-7}\cmidrule(lr){8-14}
Basis & $J^P$ & State & Channel & $M_{\rm th}$ & $\Delta_{\rm SR}$ & $\Delta_{\rm NR}$ &
Basis & $J^P$ & State & Channel & $M_{\rm th}$ & $\Delta_{\rm SR}$ & $\Delta_{\rm NR}$\\
\midrule
\endfirsthead
\multicolumn{14}{c}{\tablename\ \thetable\ -- continued from previous page}\\
\toprule
\multicolumn{7}{c}{$\bar{\mathbf3}_c\otimes\mathbf3_c$} & \multicolumn{7}{c}{$\mathbf6_c\otimes\bar{\mathbf6}_c$}\\
\cmidrule(lr){1-7}\cmidrule(lr){8-14}
Basis & $J^P$ & State & Channel & $M_{\rm th}$ & $\Delta_{\rm SR}$ & $\Delta_{\rm NR}$ &
Basis & $J^P$ & State & Channel & $M_{\rm th}$ & $\Delta_{\rm SR}$ & $\Delta_{\rm NR}$\\
\midrule
\endhead
\midrule
\multicolumn{14}{r}{Continued on next page}\\
\endfoot
\bottomrule
\endlastfoot
$(0,1)$ & $1^+$ & $1\,{}^3S_1$ & $D_s^{*+}\eta$ & $2660.1$ & $+429.1$ & $+259.6$ & $(0,0)$ & $0^+$ & $1\,{}^1S_0$ & $D_s^+\eta$ & $2516.2$ & $+1883.0$ & $+1536.2$ \\
$(1,1)$ & $0^+$ & $1\,{}^1S_0$ & $D_s^+\eta$ & $2516.2$ & $+434.9$ & $+201.7$ & $(1,0)$ & $1^+$ & $1\,{}^3S_1$ & $D_s^{*+}\eta$ & $2660.1$ & $+1733.8$ & $+1389.7$ \\
$(1,1)$ & $1^+$ & $1\,{}^3S_1$ & $D_s^{*+}\eta$ & $2660.1$ & $+372.6$ & $+167.7$ & & & & & & & \\
$(1,1)$ & $2^+$ & $1\,{}^5S_2$ & $D_s^{*+}\phi$ & $3131.7$ & $+64.1$ & $-84.3$ & & & & & & & \\
$(0,1)$ & $0^-$ & $1\,{}^3P_0$ & $D_s^{*+}\eta$ & $2660.1$ & $+846.3$ & $+656.0$ & $(0,0)$ & $1^-$ & $1\,{}^1P_1$ & $D_s^+\eta$ & $2516.2$ & $+1913.3$ & $+1570.2$ \\
$(0,1)$ & $1^-$ & $1\,{}^3P_1$ & $D_s^{*+}\eta$ & $2660.1$ & $+883.4$ & $+693.5$ & $(1,0)$ & $0^-$ & $1\,{}^3P_0$ & $D_s^{*+}\eta$ & $2660.1$ & $+1774.9$ & $+1436.8$ \\
$(0,1)$ & $2^-$ & $1\,{}^3P_2$ & $D_s^{*+}\eta$ & $2660.1$ & $+957.8$ & $+768.5$ & $(1,0)$ & $1^-$ & $1\,{}^3P_1$ & $D_s^{*+}\eta$ & $2660.1$ & $+1769.5$ & $+1430.3$ \\
$(1,1)$ & $1^-$ & $1\,{}^1P_1$ & $D_s^+\eta$ & $2516.2$ & $+1071.3$ & $+876.6$ & $(1,0)$ & $2^-$ & $1\,{}^3P_2$ & $D_s^{*+}\eta$ & $2660.1$ & $+1758.8$ & $+1417.3$ \\
$(1,1)$ & $0^-$ & $1\,{}^3P_0$ & $D_s^{*+}\eta$ & $2660.1$ & $+802.7$ & $+604.1$ & & & & & & & \\
$(1,1)$ & $1^-$ & $1\,{}^3P_1$ & $D_s^{*+}\eta$ & $2660.1$ & $+929.9$ & $+734.6$ & & & & & & & \\
$(1,1)$ & $2^-$ & $1\,{}^3P_2$ & $D_s^{*+}\eta$ & $2660.1$ & $+967.9$ & $+772.1$ & & & & & & & \\
$(1,1)$ & $1^-$ & $1\,{}^5P_1$ & $D_s^{*+}\phi$ & $3131.7$ & $+330.8$ & $+130.1$ & & & & & & & \\
$(1,1)$ & $2^-$ & $1\,{}^5P_2$ & $D_s^{*+}\phi$ & $3131.7$ & $+489.1$ & $+291.9$ & & & & & & & \\
$(1,1)$ & $3^-$ & $1\,{}^5P_3$ & $D_s^{*+}\phi$ & $3131.7$ & $+546.1$ & $+348.1$ & & & & & & & \\
$(0,1)$ & $1^+$ & $1\,{}^3D_1$ & $D_s^{*+}\eta$ & $2660.1$ & $+1169.8$ & $+986.3$ & $(0,0)$ & $2^+$ & $1\,{}^1D_2$ & $D_s^+\eta$ & $2516.2$ & $+1962.1$ & $+1624.4$ \\
$(0,1)$ & $2^+$ & $1\,{}^3D_2$ & $D_s^{*+}\eta$ & $2660.1$ & $+1179.9$ & $+996.0$ & $(1,0)$ & $1^+$ & $1\,{}^3D_1$ & $D_s^{*+}\eta$ & $2660.1$ & $+1824.8$ & $+1492.0$ \\
$(0,1)$ & $3^+$ & $1\,{}^3D_3$ & $D_s^{*+}\eta$ & $2660.1$ & $+1195.1$ & $+1010.5$ & $(1,0)$ & $2^+$ & $1\,{}^3D_2$ & $D_s^{*+}\eta$ & $2660.1$ & $+1816.9$ & $+1482.7$ \\
$(1,1)$ & $2^+$ & $1\,{}^1D_2$ & $D_s^+\eta$ & $2516.2$ & $+1352.3$ & $+1161.6$ & $(1,0)$ & $3^+$ & $1\,{}^3D_3$ & $D_s^{*+}\eta$ & $2660.1$ & $+1805.1$ & $+1468.7$ \\
$(1,1)$ & $1^+$ & $1\,{}^3D_1$ & $D_s^{*+}\eta$ & $2660.1$ & $+1189.1$ & $+998.3$ & & & & & & & \\
$(1,1)$ & $2^+$ & $1\,{}^3D_2$ & $D_s^{*+}\eta$ & $2660.1$ & $+1207.5$ & $+1016.7$ & & & & & & & \\
$(1,1)$ & $3^+$ & $1\,{}^3D_3$ & $D_s^{*+}\eta$ & $2660.1$ & $+1217.5$ & $+1025.8$ & & & & & & & \\
$(1,1)$ & $0^+$ & $1\,{}^5D_0$ & $D_s^{*+}\phi$ & $3131.7$ & $+699.8$ & $+508.5$ & & & & & & & \\
$(1,1)$ & $1^+$ & $1\,{}^5D_1$ & $D_s^{*+}\phi$ & $3131.7$ & $+711.3$ & $+520.2$ & & & & & & & \\
$(1,1)$ & $2^+$ & $1\,{}^5D_2$ & $D_s^{*+}\phi$ & $3131.7$ & $+729.6$ & $+538.5$ & & & & & & & \\
$(1,1)$ & $3^+$ & $1\,{}^5D_3$ & $D_s^{*+}\phi$ & $3131.7$ & $+748.0$ & $+556.5$ & & & & & & & \\
$(1,1)$ & $4^+$ & $1\,{}^5D_4$ & $D_s^{*+}\phi$ & $3131.7$ & $+756.1$ & $+562.9$ & & & & & & &

\end{longtable}
\endgroup

The threshold table sharpens the interpretation of the mass spectrum. Most Pauli-allowed $n=1$ states lie above their lowest spin-compatible rearrangement threshold in both SR and NR. The clearest exception is not a new state but a change of kinematic status: the antitriplet axial--axial $2^+$ basis lies above $D_s^*\phi$ in SR and below it in NR. This open-versus-closed switch is more physically discriminating than a modest numerical shift of an isolated mass because it changes the availability of the leading fall-apart channel. All displayed $P$- and $D$-wave ground orbital states remain above their assigned thresholds; their eventual widths will therefore depend on angular-momentum barriers and channel couplings rather than on kinematic closure alone. The detailed decay consequences are developed in Secs.~\ref{sec:decays} and~\ref{subsec:results_tetraquark}.

\begin{figure}[!htbp]
\centering
\includegraphics[width=0.98\textwidth]{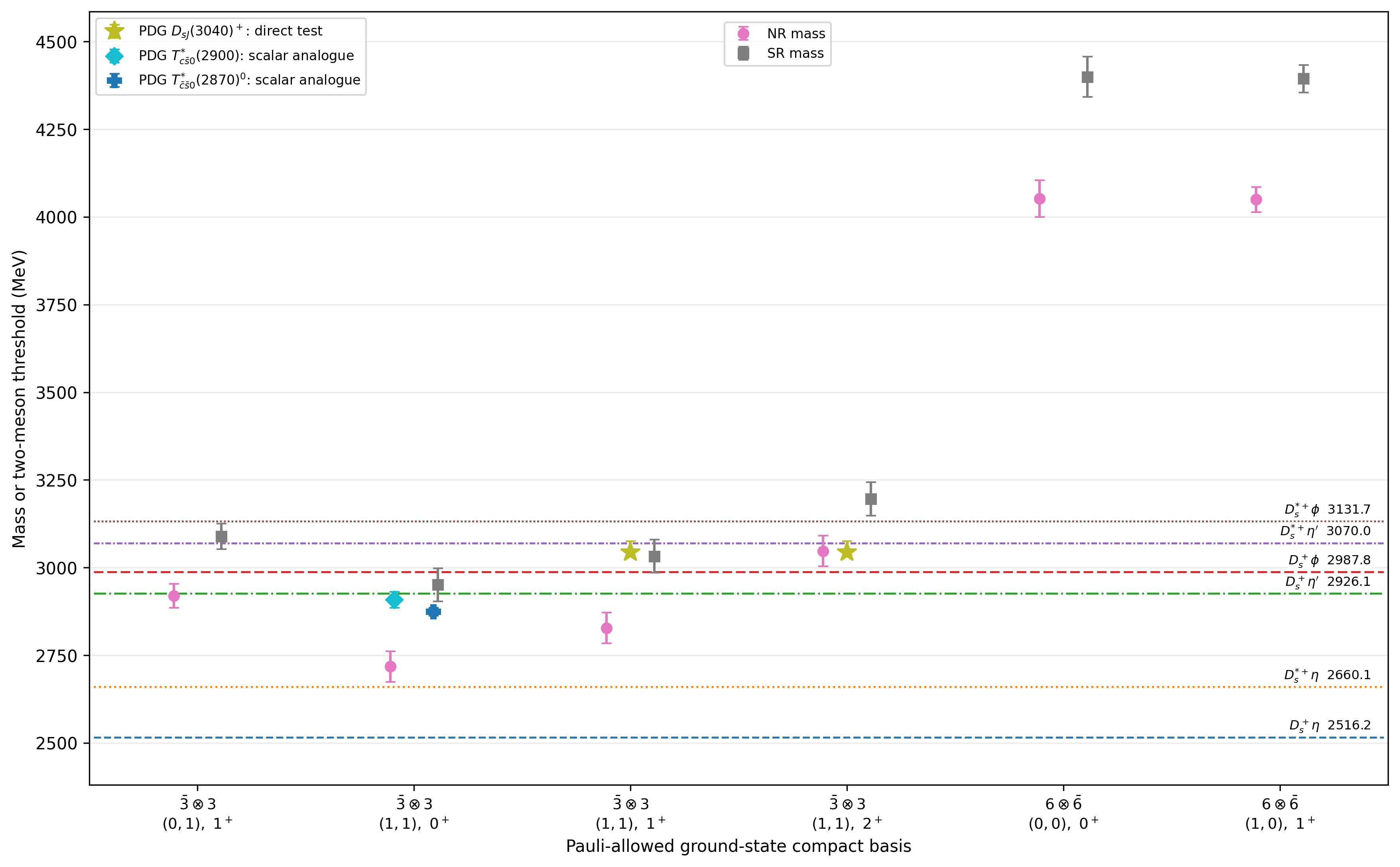}
\caption{Pauli-allowed ground-state compact-basis masses and the leading physical $D_s^{(*)}\eta^{(\prime)}$ and $D_s^{(*)}\phi$ thresholds. Experimental markers are included for mass-scale orientation; their flavor and quantum-number implications are assessed in Sec.~\ref{subsec:tq-candidate-audit}.}
\label{fig:threshold-map}
\end{figure}

Figure~\ref{fig:threshold-map} summarizes this threshold-first view by placing the compact ground-state masses and the principal $D_s^{(*)}\eta^{(\prime)}$ and $D_s^{(*)}\phi$ onsets on the same scale. Experimental markers are included only as orientation; a mass coincidence is not used as a state assignment without compatible flavor, quantum numbers, and decay information.

\section{Regge trajectories}
\label{sec:4}

Regge trajectories organize hadron spectra by relating orbital or radial quantum numbers to squared masses. This phenomenology originates in the Regge-pole and Chew--Frautschi constructions and remains widely used in spectroscopy~\cite{Regge:1959mz,Chew:1962eu,Collins:1977jy,Tang:2000tb,Selem:2006nd}. Oudichhya and collaborators have recently applied quasi-linear Regge systematics to kaon and strangeonium mesons, open-charm $D$ mesons, singly charmed baryons, and singly bottom or doubly heavy baryons~\cite{Oudichhya:2023kaon,Oudichhya:2024dmeson,Oudichhya:2023charmedbaryons,Oudichhya:2024bottombaryons}. Related analyses of fully heavy, hidden-flavor, all-light, and strange tetraquark multiplets provide useful comparative slopes and ordering patterns for the present open-charm multi-strange system~\cite{Patel:2025rsf,Patel:2025bmc,Patel:2025bdu}. We parameterize orbital trajectories in the $(L,M^2)$ plane as
\begin{equation}
    L = \alpha \, M^2 + \alpha_0,
\end{equation}
while the radial trajectories are written in terms of $n_r=n-1$ as
\begin{equation}
    n_r = \beta \, M^2 + \beta_0.
\label{eq:radial-regge}
\end{equation}
Here $\alpha$ and $\beta$ are the orbital and radial slopes. Exact linearity is not expected for heavy-light $D_s$ mesons or compact $cs\bar{s}\bar{s}$ tetraquarks because unequal constituent masses, spin-dependent interactions, and the internal diquark--antidiquark structure distort the spectrum~\cite{Tang:2000tb,Selem:2006nd,Oudichhya:2024dmeson,Patel:2025rsf}. Nevertheless, the fitted trajectories provide a compact measure of spectral regularity and facilitate direct comparison between the SR and NR treatments.

Tables~\ref{tab:regge_meson} and~\ref{tab:regge_tetraquark} summarize the orbital and radial fits for the $D_s$ and $cs\bar s\bar s$ sectors, with the two sets of parameters displayed side by side for direct comparison.

\begin{table}[!htbp]
\centering
\caption{Orbital and radial Regge slopes for the $D_s$ meson family, displayed side by side. The ratios are SR/NR, and all slopes are in GeV$^{-2}$.}
\label{tab:regge_meson}
\tablefont
\setlength{\tabcolsep}{2.4pt}
\begin{tabular}{@{}ccccc@{\hspace{8pt}}ccccc@{}}
\toprule
\multicolumn{5}{c}{Orbital: $L=\alpha M^2+\alpha_0$} & \multicolumn{5}{c}{Radial: $n_r=\beta M^2+\beta_0$}\\
\cmidrule(lr){1-5}\cmidrule(lr){6-10}
$S$ & fixed $n$ & $\alpha_{\rm SR}$ & $\alpha_{\rm NR}$ & $R_\alpha$ & $S$ & fixed $L$ & $\beta_{\rm SR}$ & $\beta_{\rm NR}$ & $R_\beta$\\
\midrule
0 & 0 & 0.6165 & 0.6088 & 1.013 & 0 & 0 & 0.4296 & 0.4197 & 1.024\\
0 & 1 & 0.7347 & 0.7145 & 1.028 & 0 & 1 & 0.4854 & 0.4700 & 1.033\\
0 & 2 & 0.7630 & 0.7577 & 1.007 & 0 & 2 & 0.4900 & 0.4836 & 1.013\\
0 & 3 & 0.7382 & 0.7800 & 0.9463 & 0 & 3 & 0.4795 & 0.4911 & 0.9764\\
1 & 0 & 0.7077 & 0.6817 & 1.038 & 1 & 0 & 0.4514 & 0.4339 & 1.040\\
1 & 1 & 0.7833 & 0.7509 & 1.043 & 1 & 1 & 0.4794 & 0.4633 & 1.035\\
1 & 2 & 0.7840 & 0.7779 & 1.008 & 1 & 2 & 0.4909 & 0.4842 & 1.014\\
1 & 3 & 0.7328 & 0.7889 & 0.9290 & 1 & 3 & 0.4800 & 0.4916 & 0.9764\\
\bottomrule
\end{tabular}
\end{table}

The tetraquark trajectories are constructed only after the Pauli projection because the sextet axial--axial $S_T=0,1,2$ families are forbidden. The radial fits use Eq.~\eqref{eq:radial-regge} with the $n=1,2,3$ $S$-wave masses, while the orbital fits use the standard spin-degeneracy-weighted centroids of the $n=1$ states at $L=0,1,2$~\cite{Selem:2006nd,Tang:2000tb},
\begin{equation}
L=\alpha M_{\rm cog}^2+\alpha_0,
\qquad
M_{\rm cog}^2=\frac{\sum_J(2J+1)M_J^2}{\sum_J(2J+1)}.
\label{eq:orbital-regge-corrected}
\end{equation}

\begin{table}[!htbp]
\centering
\caption{Pauli-filtered tetraquark Regge parameters, with radial and orbital fits displayed side by side. Slopes are in GeV$^{-2}$.}
\label{tab:regge_tetraquark}
\tablefont
\setlength{\tabcolsep}{2.1pt}
\begin{tabular}{@{}L{43mm}C{9mm}C{14mm}C{17mm}C{12mm}C{14mm}C{17mm}C{12mm}@{}}
\toprule
& & \multicolumn{3}{c}{Radial} & \multicolumn{3}{c}{Orbital}\\
\cmidrule(lr){3-5}\cmidrule(lr){6-8}
Pauli-allowed family & Scheme & Slope & Intercept & $R^2$ & Slope & Intercept & $R^2$\\
\midrule
$\bar{\mathbf3}\otimes\mathbf3\ (0,1),\ S_T=1$ & SR & $0.2847$ & $-2.7648$ & $0.9901$ & $0.3737$ & $-3.6276$ & $0.9793$ \\
$\bar{\mathbf3}\otimes\mathbf3\ (0,1),\ S_T=1$ & NR & $0.2968$ & $-2.5714$ & $0.9931$ & $0.4035$ & $-3.4957$ & $0.9841$ \\
$\bar{\mathbf3}\otimes\mathbf3\ (1,1),\ S_T=0$ & SR & $0.2553$ & $-2.2798$ & $0.9839$ & $0.3085$ & $-2.7575$ & $0.9649$ \\
$\bar{\mathbf3}\otimes\mathbf3\ (1,1),\ S_T=0$ & NR & $0.2625$ & $-1.9914$ & $0.9869$ & $0.3135$ & $-2.3878$ & $0.9622$ \\
$\bar{\mathbf3}\otimes\mathbf3\ (1,1),\ S_T=1$ & SR & $0.2681$ & $-2.5183$ & $0.9869$ & $0.3367$ & $-3.1647$ & $0.9713$ \\
$\bar{\mathbf3}\otimes\mathbf3\ (1,1),\ S_T=1$ & NR & $0.2775$ & $-2.2668$ & $0.9899$ & $0.3518$ & $-2.8801$ & $0.9722$ \\
$\bar{\mathbf3}\otimes\mathbf3\ (1,1),\ S_T=2$ & SR & $0.2995$ & $-3.1002$ & $0.9930$ & $0.4126$ & $-4.2668$ & $0.9870$ \\
$\bar{\mathbf3}\otimes\mathbf3\ (1,1),\ S_T=2$ & NR & $0.3165$ & $-2.9720$ & $0.9960$ & $0.4670$ & $-4.3731$ & $0.9947$ \\
$\mathbf6\otimes\bar{\mathbf6}\ (0,0),\ S_T=0$ & SR & $0.7912$ & $-15.3175$ & $0.9999$ & $2.7944$ & $-53.9821$ & $0.9814$ \\
$\mathbf6\otimes\bar{\mathbf6}\ (0,0),\ S_T=0$ & NR & $0.8007$ & $-13.1532$ & $0.9999$ & $2.7173$ & $-44.5284$ & $0.9821$ \\
$\mathbf6\otimes\bar{\mathbf6}\ (1,0),\ S_T=1$ & SR & $0.7918$ & $-15.2928$ & $0.9999$ & $2.7955$ & $-53.8758$ & $0.9814$ \\
$\mathbf6\otimes\bar{\mathbf6}\ (1,0),\ S_T=1$ & NR & $0.8010$ & $-13.1422$ & $0.9999$ & $2.7178$ & $-44.4807$ & $0.9821$
\\
\bottomrule
\end{tabular}
\end{table}

The fitted parameters are listed in Tables~\ref{tab:regge_meson} and~\ref{tab:regge_tetraquark}. Their role is deliberately limited: near-parallel SR and NR meson slopes test whether the model preserves global ordering, while the tetraquark slopes summarize the calculated spacing after Pauli projection. Neither linearity nor a particular slope constitutes independent evidence that an observed resonance is compact; Regge systematics are used here only as phenomenological organization, as in related meson and tetraquark studies~\cite{Selem:2006nd,Oudichhya:2023kaon,Oudichhya:2024dmeson,Patel:2025rsf,Patel:2025bmc,Patel:2025bdu}. In particular, the unusually steep sextet orbital slopes arise from the compressed heavy sextet spectrum and are best regarded as model-specific diagnostics. Further interpretation is given in Secs.~\ref{subsec:results_ds_meson} and~\ref{subsec:results_tetraquark}.

The detailed $D_s$ orbital and radial Regge plots are provided as Supplemental Figs.~S3 and S4. In the main text we retain the fitted parameters because the relevant conclusion is quantitative: SR and NR give similar global slopes even when individual excited masses differ. These fits are diagnostics of spectral regularity and are not independent evidence for a tetraquark interpretation.

\begin{figure}[!htbp]
\centering
\begin{subfigure}[t]{0.49\textwidth}
\centering
\includegraphics[width=\linewidth]{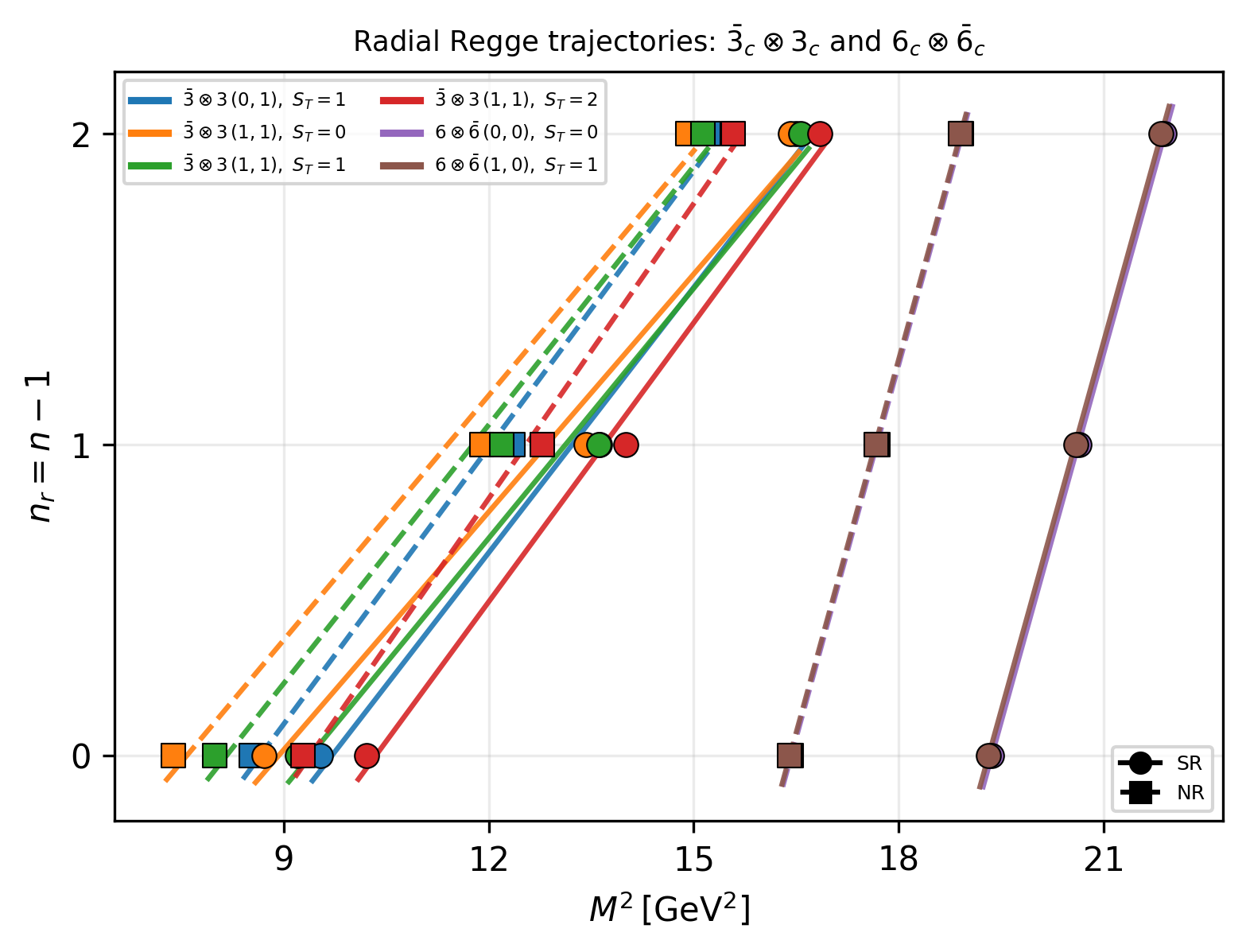}
\caption{Radial trajectories in the $(n_r,M^2)$ plane.}
\label{fig:regge-radial-corrected}
\end{subfigure}\hfill
\begin{subfigure}[t]{0.49\textwidth}
\centering
\includegraphics[width=\linewidth]{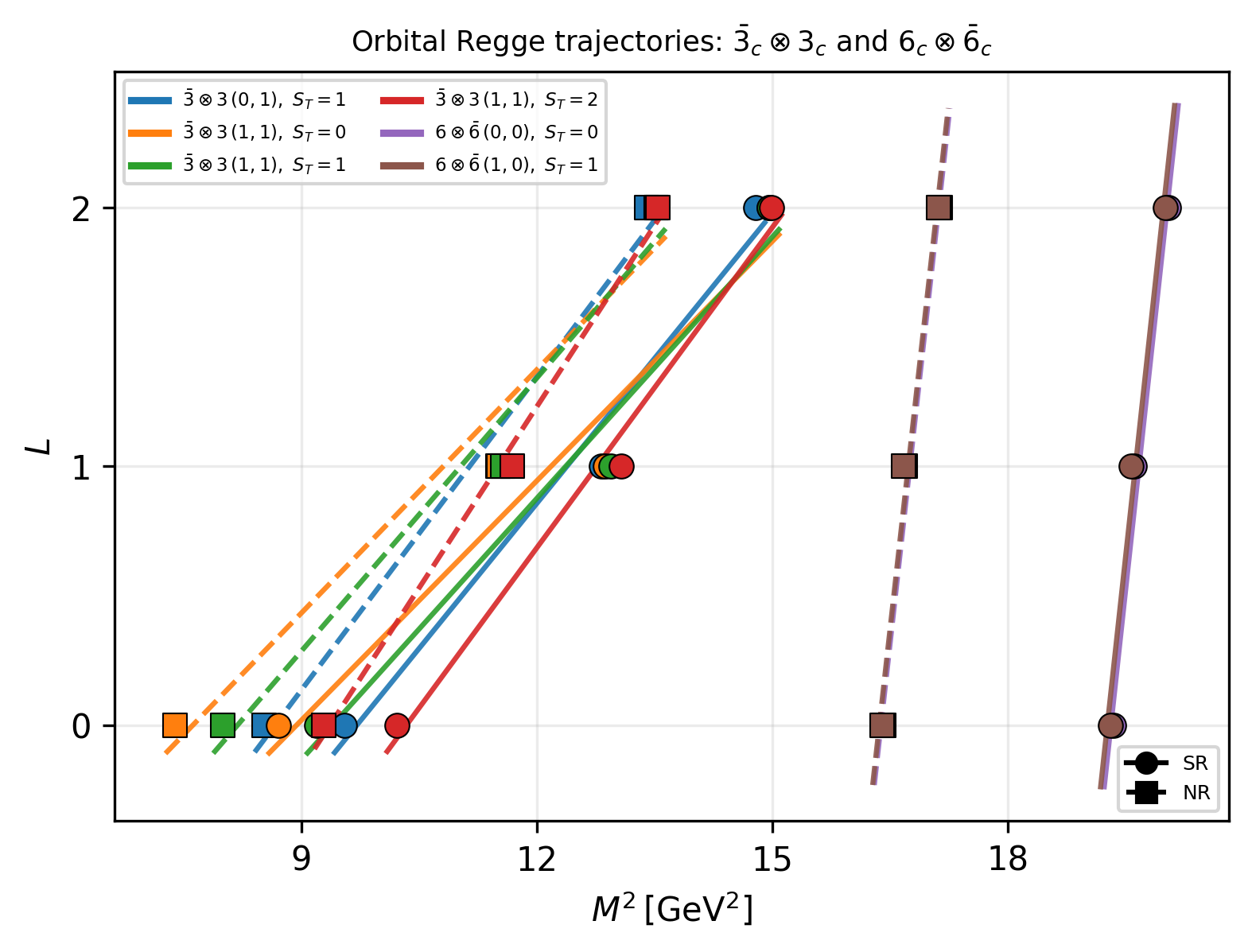}
\caption{Orbital trajectories in the $(L,M^2)$ plane.}
\label{fig:regge-orbital-corrected}
\end{subfigure}
\caption{Radial and orbital Regge trajectories of the Pauli-allowed $\bar{\mathbf3}_c\otimes\mathbf3_c$ and $\mathbf6_c\otimes\bar{\mathbf6}_c$ tetraquark families.}
\end{figure}
\FloatBarrier

\section{Decays}
\label{sec:decays}
\label{sec:decay}

Decay observables test channel accessibility and internal structure~\cite{ParticleDataGroup:2026rpp,Guo:2017jvc}. Hidden-strange and weak-$D_s$ inputs are established first, followed by the tetraquark rearrangement analysis based on Fierz-compatible fall-apart topologies~\cite{Chen:2020fierz}, exact color--spin projectors, and phase space. The current-specific requirements for absolute strong couplings are given in Sec.~\ref{subsec:three-point-requirements}.

\subsection{Meson decays: hidden-strange benchmark}

The meson calculations supply the daughter branching fractions used in cascade reconstruction and benchmark the hidden-strange sector.

Hidden-strange mesons enter directly into the lowest rearrangement channels of the compact $cs\bar{s}\bar{s}$ system. The $\phi(1020)$ offers the cleanest $s\bar{s}$ benchmark because its leptonic, strong, and radiative modes can be expressed within the standard local two-point and three-point QCD-sum-rule framework~\cite{Shifman:1978bx,Shifman:1978by,DeFazioPennington2001,DeFazioPennington2000}. We employ the currents
\begin{equation}
 j_\mu^\phi = \bar s \gamma_\mu s, \qquad j_5^s = \bar s i\gamma_5 s, \qquad J_s = \bar s s, \qquad j_5^K = \bar q i\gamma_5 s,
\end{equation}
with the matrix-element definitions
\begin{equation}
\langle 0|j_\mu^\phi|\phi(q,\epsilon)\rangle = f_\phi m_\phi \epsilon_\mu, \qquad
\langle 0|j_5^s|\eta\rangle = A, \qquad
\langle 0|j_5^s|\eta'\rangle = A', \qquad
\langle 0|J_s|f_0\rangle = m_{f_0}\tilde f_{f_0}.
\end{equation}
The two-point QCD-sum-rule benchmark inputs used in the decay calculation follow the hidden-strange sum-rule analyses of Refs.~\cite{DeFazioPennington2001,DeFazioPennington2000}:
\begin{equation}
 m_\phi = 1.023\pm0.013~\mathrm{GeV}, \qquad f_\phi = 0.240\pm0.005~\mathrm{GeV},
\end{equation}
\begin{equation}
 A = 0.115\pm0.004~\mathrm{GeV}^2, \qquad A' = 0.153\pm0.011~\mathrm{GeV}^2, \qquad \tilde f_{f_0} = 0.180\pm0.015~\mathrm{GeV}.
\end{equation}
The strong and radiative channels are normalized with representative QCD-sum-rule/phenomenological couplings~\cite{Aliev:2003phiKK,DeFazioPennington2001,DeFazioPennington2000}:
\begin{equation}
\langle K(p)\bar K(p')|\phi(q,\epsilon)\rangle = g_{\phi KK}(q^2)\,\epsilon\cdot (p-p'), \qquad g_{\phi KK}=4.82\pm1.30,
\end{equation}
\begin{align}
 g_{\phi\eta\gamma}&=0.64\pm0.06~\mathrm{GeV}^{-1}, &
 g_{\phi\eta'\gamma}&=0.98\pm0.20~\mathrm{GeV}^{-1},\nonumber\\
 g_{\phi\pi\gamma}&=0.125\pm0.004, &
 F_1^{\phi f_0}(0)&=0.34\pm0.07~\mathrm{GeV}^{-1}.
\end{align}
The associated benchmark widths are calculated from
\begin{equation}
\Gamma(\phi\to \ell^+\ell^-)=
\frac{4\pi\alpha_{\rm em}^2Q_s^2}{3m_\phi}f_\phi^2
\left(1+\frac{2m_\ell^2}{m_\phi^2}\right)
\sqrt{1-\frac{4m_\ell^2}{m_\phi^2}},
\qquad Q_s=-\frac{1}{3},
\end{equation}
\begin{equation}
\Gamma(\phi\to K\bar K)=\frac{g_{\phi KK}^2}{6\pi m_\phi^2}|\vec p_K|^3,
\qquad
|\vec p_K|=\frac{\lambda^{1/2}(m_\phi^2,m_K^2,m_{\bar K}^2)}{2m_\phi},
\end{equation}
\begin{equation}
\Gamma(\phi\to P\gamma)=\frac{\alpha_{\rm em}g_{\phi P\gamma}^2}{24}
\left(\frac{m_\phi^2-m_P^2}{m_\phi}\right)^3,
\qquad P=\eta,\eta',
\end{equation}
and
\begin{equation}
\Gamma(\phi\to f_0\gamma)=
\frac{\alpha_{\rm em}[F_1^{\phi f_0}(0)]^2(m_\phi^2-m_{f_0}^2)(m_\phi^2+m_{f_0}^2)^2}{216m_\phi^3}.
\end{equation}
Table~\ref{tab:ss_phi_decay} summarizes the resulting widths and branching fractions used in the subsequent cascade construction.

\begin{table}[!htbp]
\centering
\caption{Present benchmark partial widths for hidden-strange $\phi(1020)$ decays compared with PDG 2026 branching fractions and the corresponding PDG partial widths~\cite{ParticleDataGroup:2026rpp}. The PDG partial widths are obtained by multiplying $\Gamma_\phi^{\rm PDG}=4.249\pm0.013~\mathrm{MeV}$ by the listed branching fractions.}
\label{tab:ss_phi_decay}
\tablefont
\begin{tabular}{@{}L{23mm}C{27mm}C{30mm}C{30mm}C{30mm}@{}}
\toprule
Mode & Present width & Present branching fraction & PDG branching fraction & PDG partial width \\
\midrule
$\phi\to e^+e^-$ & $1.40\pm0.06~\mathrm{keV}$ & $(3.29\pm0.14)\times10^{-4}$ & $(2.979\pm0.033)\times10^{-4}$ & $1.266\pm0.015~\mathrm{keV}$ \\
\addlinespace
$\phi\to \mu^+\mu^-$ & $1.40\pm0.06~\mathrm{keV}$ & $(3.29\pm0.14)\times10^{-4}$ & $(2.85\pm0.22)\times10^{-4}$ & $1.211\pm0.094~\mathrm{keV}$ \\
\addlinespace
$\phi\to K^+K^-$ & $2.42\pm1.31~\mathrm{MeV}$ & $0.570\pm0.308$ & $0.491\pm0.005$ & $2.086\pm0.022~\mathrm{MeV}$ \\
\addlinespace
$\phi\to K^0\bar K^0$ & $1.60\pm0.86~\mathrm{MeV}$ & $0.377\pm0.202$ & $0.339\pm0.004$ & $1.440\pm0.018~\mathrm{MeV}$ \\
\addlinespace
$\phi\to \eta\gamma$ & $47.5\pm8.9~\mathrm{keV}$ & $(1.13\pm0.21)\times10^{-2}$ & $(1.301\pm0.024)\times10^{-2}$ & $55.28\pm1.03~\mathrm{keV}$ \\
\addlinespace
$\phi\to \eta'\gamma$ & $0.50\pm0.21~\mathrm{keV}$ & $(1.18\pm0.49)\times10^{-4}$ & $(6.21\pm0.20)\times10^{-5}$ & $0.264\pm0.009~\mathrm{keV}$ \\
\addlinespace
$\phi\to \pi^0\gamma$ & $4.59\pm0.29~\mathrm{keV}$ & $(1.08\pm0.07)\times10^{-3}$ & $(1.32\pm0.05)\times10^{-3}$ & $5.61\pm0.21~\mathrm{keV}$ \\
\addlinespace
$\phi\to f_0(980)\gamma$ & $1.16\pm0.48~\mathrm{keV}$ & $(2.73\pm1.13)\times10^{-4}$ & $(3.22\pm0.19)\times10^{-4}$ & $1.37\pm0.08~\mathrm{keV}$ \\
\bottomrule
\end{tabular}
\end{table}

\subsection{Weak decays of the \texorpdfstring{$D_s$}{Ds} meson}

The $D_s^+$ decay analysis includes charm decay with a spectator $\bar s$, weak annihilation, and the highly suppressed spectator-$c$ transition $\bar s\to\bar u$. The corresponding semileptonic and factorized nonleptonic interactions follow the standard effective-Hamiltonian and factorization treatment~\cite{Bauer:1987BSW,Yu:2023galkin}:
\begin{align}
\mathcal H_{\rm sl}&=\frac{G_F}{\sqrt2}V_{cq}^{*}\,[\bar q\gamma_\mu(1-\gamma_5)c]\,[\bar\nu_\ell\gamma^\mu(1-\gamma_5)\ell],\\
H_{\rm nl}&=\frac{G_F}{\sqrt2}V_{cq_1}^{*}V_{uq_2}\,[c_1O_1+c_2O_2],
\qquad a_1=c_1+\frac{c_2}{N_c}=1.09,
\end{align}
where $c_1=1.26$, $c_2=-0.51$, and $N_c=3$~\cite{Yu:2023galkin}. The $D_s\to X$ form factors are represented by the standard three-point QCD correlator~\cite{Shifman:1978bx,Shifman:1978by,Bediaga:2003zh,Du:2003DsPhi}
\begin{equation}
\Pi_{\{\mu\cdots\}}(p,p')=i^2\!\int d^4x\,d^4y\,e^{ip'\cdot x}e^{i(p-p')\cdot y}
\langle0|T\{J_X(x)J_W(y)J_{D_s}^{\dagger}(0)\}|0\rangle,
\end{equation}
with $J_{D_s}=\bar s i\gamma_5c$, $q=p-p'$, and $P=p+p'$. Pseudoscalar and scalar transitions use
\begin{align}
\langle P(p')|\bar q\gamma_\mu c|D_s(p)\rangle&=f_+(q^2)P_\mu+f_-(q^2)q_\mu,\\
f_0(q^2)&=f_+(q^2)+\frac{q^2}{m_{D_s}^2-m_P^2}f_-(q^2),
\qquad F_{1,0}\equiv f_{+,0},
\end{align}
whereas vector daughters are described by $V$, $A_0$, $A_1$, and $A_2$ with the standard condition $A_0(0)=A_3(0)$. We use $V^{D_s\to\phi}(0)=1.21$, $A_0(0)=0.42$, $A_1(0)=0.55$, and $A_2(0)=0.59$ from the QCD-sum-rule analysis of Du, Li, and Yang~\cite{Du:2003DsPhi}, together with
\begin{equation}
f_+^{D_s\to\eta}(0)=0.495,\qquad
f_+^{D_s\to\eta'}(0)=0.558,\qquad
F_1^{D_s\to f_0}(0)=0.516.
\end{equation}
The $D_s\to\eta^{(\prime)}$ values are the central light-cone sum-rule results of Duplancic and Melic~\cite{Duplancic:2015eta}; the $D_s\to f_0$ transition is treated within the QCD-sum-rule framework discussed for $D_s\to\phi,f_0$ decays in Ref.~\cite{Bediaga:2003zh}.
When a channel-specific shape is unavailable, $F(q^2)=F(0)/(1-q^2/m_{\rm pole}^2)$ is used.

\subsection{Integrated semileptonic benchmark}

The channel-specific form factors are used only as an internal benchmark for the daughter-decay weights. Detailed form-factor and differential-rate panels are moved to the Supplemental Material to keep the phenomenological discussion focused.

For light charged leptons, the principal differential rates are written in the standard form-factor/helicity representation~\cite{Du:2003DsPhi,Duplancic:2015eta,Bediaga:2003zh}:
\begin{align}
\frac{d\Gamma(D_s\to P\ell\nu)}{dq^2}&=\frac{G_F^2|V_{cq}|^2}{24\pi^3}|\mathbf p_P|^3|f_+(q^2)|^2,\\
\frac{d\Gamma(D_s\to S\ell\nu)}{dq^2}&=\frac{G_F^2|V_{cq}|^2}{24\pi^3}|\mathbf p_S|^3|F_1(q^2)|^2,\\
\frac{d\Gamma(D_s\to V\ell\nu)}{dq^2}&=\frac{G_F^2|V_{cq}|^2}{96\pi^3m_{D_s}^2}q^2|\mathbf p_V|
\bigl(|H_+|^2+|H_-|^2+|H_0|^2\bigr),
\end{align}
where $|\mathbf p_X|=\lambda^{1/2}(m_{D_s}^2,m_X^2,q^2)/(2m_{D_s})$ and the helicity amplitudes have their standard $V,A_1,A_2$ definitions. Integration over $m_\ell^2\le q^2\le(m_{D_s}-m_X)^2$ gives $\mathcal B=\Gamma\tau_{D_s}$. The numerical form-factor profiles and electron--muon differential distributions are provided in Supplemental Figs.~S1 and S2; the main text retains only the integrated observables used in the comparison below.

For factorized $D_s\to M_1M_2$ modes, we use the conventional external-emission factorization form~\cite{Bauer:1987BSW,Yu:2023galkin,Bediaga:2003zh}:
\begin{align}
\mathcal A(D_s\to P_1P_2)&=\frac{G_F}{\sqrt2}V_{cq_1}^{*}V_{uq_2}a_1 f_{P_2}(m_{D_s}^2-m_{P_1}^2)f_0^{D_s\to P_1}(m_{P_2}^2),\\
\Gamma(D_s\to M_1M_2)&=\frac{|\mathbf p|}{8\pi m_{D_s}^2}|\mathcal A|^2.
\end{align}
The spectator-$c$ channel has $q_{\max}^2(D_s\to D^0)=1.0714\times10^{-2}$~GeV$^2$ and
\begin{equation}
\mathcal B(D_s^+\to D^0e^+\nu_e)=3.07\times10^{-8}|f_+^{D_s\to D}(0)|^2,
\end{equation}
which explains its severe phase-space suppression. To keep this validation layer proportionate to its role, Table~\ref{tab:ds_decay_combined} retains only representative experimentally constrained modes. The complete nonzero semileptonic, spectator-$c$, and factorized nonleptonic inventory, including product branching fractions and limits, is moved to Supplemental Table~S6. The compact main-text table is sufficient to show the relevant conclusion: semileptonic agreement is generally better than the channel-dependent factorized nonleptonic description, so these calculations are used to benchmark daughter weights rather than as a precision weak-decay fit.

\begin{table}[!htbp]
\centering
\caption{Representative $D_s^+$ decay benchmarks retained in the main text. The full set of calculated nonzero semileptonic, spectator-$c$, and factorized nonleptonic modes, including limits and product branching fractions, is given in Supplemental Table~S6. Branching fractions are dimensionless.}
\label{tab:ds_decay_combined}
\tablefont
\setlength{\tabcolsep}{4.0pt}
\begin{tabular}{@{}L{48mm}C{29mm}C{29mm}C{29mm}@{}}
\toprule
Channel & PDG 2026 & SR & NR \\
\midrule
$D_s^+\to \eta\,e^+\nu_e$ & $(2.27\pm0.06)\times10^{-2}$ & $2.502\times10^{-2}$ & $2.624\times10^{-2}$ \\
$D_s^+\to \eta'\,e^+\nu_e$ & $(8.1\pm0.4)\times10^{-3}$ & $4.815\times10^{-3}$ & $5.123\times10^{-3}$ \\
$D_s^+\to \phi\,e^+\nu_e$ & $(2.34\pm0.12)\times10^{-2}$ & $2.564\times10^{-2}$ & $2.745\times10^{-2}$ \\
$D_s^+\to K^{0}\,e^+\nu_e$ & $(2.88\pm0.26)\times10^{-3}$ & $2.615\times10^{-3}$ & $2.742\times10^{-3}$ \\
\midrule
$D_s^+\to \eta\,\pi^+$ & $(1.686\pm0.027)\times10^{-2}$ & $3.796\times10^{-2}$ & $3.887\times10^{-2}$ \\
$D_s^+\to \eta\,\rho^+$ & $(8.9\pm0.8)\times10^{-2}$ & $7.059\times10^{-2}$ & $7.325\times10^{-2}$ \\
$D_s^+\to \eta'\,\rho^+$ & $(5.8\pm1.5)\times10^{-2}$ & $1.328\times10^{-2}$ & $1.451\times10^{-2}$ \\
$D_s^+\to \phi\,\pi^+$ & $(4.5\pm0.4)\times10^{-2}$ & $3.336\times10^{-2}$ & $3.458\times10^{-2}$ \\
$D_s^+\to \phi\,\rho^+$ & $(5.59\pm0.34)\times10^{-2}$ & $1.261\times10^{-1}$ & $1.329\times10^{-1}$ \\
\bottomrule
\end{tabular}
\end{table}

Table~\ref{tab:ds_decay_combined} shows the benchmark pattern used to judge the daughter-decay inputs; the full numerical inventory is retained in Supplemental Table~S6. Section~\ref{subsec:results_ds_meson} evaluates the experimentally constrained modes and their limitations.

\subsection{Tetraquark rearrangement decays}
\label{subsec:tq_decay_framework}

\begin{figure}[!htbp]
\centering
\includegraphics[width=0.98\textwidth]{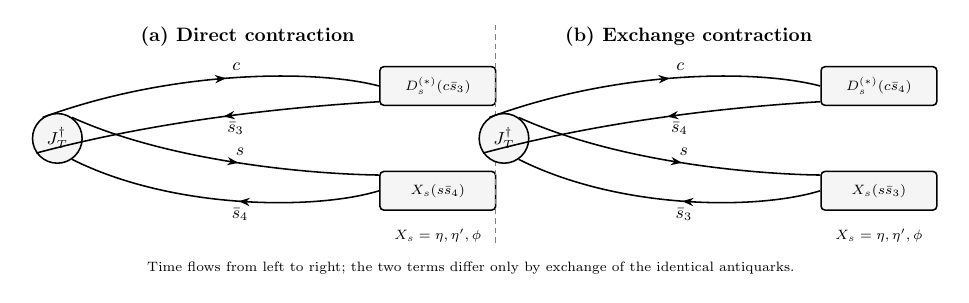}
\caption{Quark-line representation of the direct and exchange contractions for $cs\bar s_3\bar s_4\to D_s^{(*)}X_s$, with $X_s=\eta,\eta',\phi$.  Four continuous constituent lines emphasize rearrangement without additional pair creation.  The two contributions differ only by interchange of the identical antiquarks and must be combined at the amplitude level.}
\label{fig:tq-direct-exchange}
\end{figure}

The leading flavor-rearrangement process is
\begin{equation}
cs\bar{s}\bar{s}\longrightarrow(c\bar{s})(s\bar{s}),
\label{eq:tq_rearrangement_topology}
\end{equation}
where $(c\bar s)_{0^-}=D_s$, $(c\bar s)_{1^-}=D_s^*$, $(s\bar s)_{0^-}\to\eta,\eta'$, and $(s\bar s)_{1^-}=\phi$. This quark-line rearrangement/fall-apart organization is the Fierz-compatible topology used in compact tetraquark decay analyses~\cite{Jaffe:1976ig,Jaffe:1976ih,Chen:2020fierz}. The two identical antiquarks generate coherent direct and exchange contractions. Figure~\ref{fig:tq-direct-exchange} distinguishes them by temporarily labeling the antiquarks $\bar s_3$ and $\bar s_4$.

For the scalar--axial current of Eq.~\eqref{eq:current-sa1}, a representative two-point Wick contraction has the schematic direct-minus-exchange form
\begin{align}
\Pi_{\mu\nu}(p)={}&i\,{\cal C}^{bc,de}{\cal C}^{b'c',d'e'}
\int d^4x\,e^{ip\cdot x}
\operatorname{Tr}\!\left[\gamma_5S_s^{cc'}(x)\gamma_5
CS_c^{T\,bb'}(x)C\right]\nonumber\\
&\times\Big\{
\operatorname{Tr}\!\left[\gamma_\mu CS_s^{T\,ee'}(-x)C\gamma_\nu S_s^{dd'}(-x)\right]
-\operatorname{Tr}\!\left[\gamma_\mu CS_s^{T\,ed'}(-x)C\gamma_\nu S_s^{de'}(-x)\right]
\Big\},
\label{eq:direct-exchange-contraction}
\end{align}
where ${\cal C}={\cal A}$ or ${\cal S}$ for the two color sectors. The relative minus sign is the identical-fermion exchange phase, and the two contractions must be combined before squaring the amplitude.

For the direct pairing $(13)(24)=(c\bar s)(s\bar s)$, the normalized color wave functions follow the standard recoupling of diquark--antidiquark color bases into meson--meson singlet/octet components~\cite{Jaffe:1976ih,Ebert:2010af,Chen:2020fierz}:
\begin{subequations}
\begin{align}
|\bar{\mathbf 3}_c\mathbf 3_c\rangle&=\frac{1}{\sqrt3}|\mathbf1\mathbf1\rangle-\sqrt{\frac23}|\mathbf8\mathbf8\rangle,\\
|\mathbf6_c\bar{\mathbf6}_c\rangle&=\sqrt{\frac23}|\mathbf1\mathbf1\rangle+\frac{1}{\sqrt3}|\mathbf8\mathbf8\rangle.
\end{align}
\label{eq:tq_color_recoupling}
\end{subequations}
The four constituent spins are recoupled from the diquark pairing $(12)(34)$ to the meson pairing $(13)(24)$, as in standard tetraquark Fierz/recoupling analyses~\cite{Jaffe:1976ih,Chen:2020fierz}. The overlap is expressed by the Wigner-$9j$ coefficient
\begin{equation}
C_{\rm spin}=\sqrt{(2S_d+1)(2S_{\bar d}+1)(2S_{13}+1)(2S_{24}+1)}
\begin{Bmatrix}
\tfrac12&\tfrac12&S_d\\
\tfrac12&\tfrac12&S_{\bar d}\\
S_{13}&S_{24}&S_T
\end{Bmatrix}.
\label{eq:tq-spin-9j}
\end{equation}
Here $P$ and $V$ denote spin-zero and spin-one mesons. Multiplying the spin overlap by the singlet--singlet color coefficient in Eq.~\eqref{eq:tq_color_recoupling} gives the full color--spin projector. The nonzero spin factors are
\begin{align}
(1,1),\ 0^+ &: \quad C_{PP}=+\frac{\sqrt3}{2},\qquad C_{VV}=-\frac12,\nonumber\\
(1,1),\ 1^+ &: \quad C_{PV}=C_{VP}=+\frac{1}{\sqrt2},\qquad C_{VV}=0,\nonumber\\
(1,1),\ 2^+ &: \quad C_{VV}=+1,\nonumber\\
(0,1),\ 1^+ &: \quad C_{PV}=+\frac12,\qquad C_{VP}=-\frac12,\qquad C_{VV}=+\frac{1}{\sqrt2},\nonumber\\
(0,0),\ 0^+ &: \quad C_{PP}=+\frac12,\qquad C_{VV}=+\frac{\sqrt3}{2},\nonumber\\
(1,0),\ 1^+ &: \quad C_{PV}=-\frac12,\qquad C_{VP}=+\frac12,\qquad C_{VV}=+\frac{1}{\sqrt2}.
\label{eq:tq-spin-recouplings-compact}
\end{align}
For the antitriplet--triplet and sextet--antisextet sectors, these spin factors are multiplied by $1/\sqrt3$ and $\sqrt{2/3}$, respectively. The physical $\eta$ and $\eta'$ amplitudes are then obtained from the strange-flavor projection in Eq.~\eqref{eq:eta-mixing-tq}.

The recoupling algebra fixes the leading channel content of each basis. The antitriplet axial--axial $0^+$ state contains $PP$ and $VV$ components; the corresponding $1^+$ state contains $PV$ and $VP$ but no $VV$ term in this limit; and the axial--axial $2^+$ state couples only to $VV$. Both mixed-spin $1^+$ bases contain $PV$, $VP$, and $VV$, so $D_s^{*+}\phi$ is a leading $S$-wave channel for $(0,1)$ and $(1,0)$. The sextet scalar--scalar $0^+$ basis has a comparatively large $VV$ projection. The strange component of the physical pseudoscalars is represented in the quark-flavor basis by
\begin{equation}
|\eta_s\rangle=-\sin\varphi_P|\eta\rangle+\cos\varphi_P|\eta'\rangle,
\qquad \varphi_P=(39.3\pm1.0)^\circ.
\label{eq:eta-mixing-tq}
\end{equation}
We use the quark-flavor-basis mixing angle determined in the Feldmann--Kroll--Stech scheme~\cite{Feldmann:1998FKS}.

\subsection{Three-point sum-rule requirements and reduced coefficients}
\label{subsec:three-point-requirements}
For a channel $T_J(p')\to H_1(p)H_2(q)$, the local three-point QCD sum rule begins with the standard three-current correlator construction~\cite{Shifman:1978bx,Shifman:1978by,Yang:2023evp,Chen:2020fierz}:
\begin{equation}
\Pi_i(p,q)=i^2\int d^4x\,d^4y\,e^{ip\cdot x+iq\cdot y}
\langle0|T\{j_{H_1}(x)j_{H_2}(y)J_T^\dagger(0)\}|0\rangle,
\label{eq:tq_3pt_rearrangement}
\end{equation}
with a hadronic pole term
\begin{equation}
\Pi_i^{\rm had}=\frac{\Lambda_iG_i}{(M_T^2-p'^2)(m_1^2-p^2)(m_2^2-q^2)}+\cdots,
\end{equation}
and a QCD representation
\begin{equation}
\Pi_i^{\rm QCD}=\int ds\,du\,\frac{\rho_i^{\rm QCD}(s,u)}{(s-p^2)(u-q^2)}+\cdots.
\end{equation}
An absolute coupling requires the current-specific double spectral density, continuum subtraction, Borel windows, pole contribution, OPE-convergence analysis, and Lorentz-structure stability for every Pauli-allowed current, which are standard consistency requirements in QCD sum-rule extractions~\cite{Shifman:1978bx,Shifman:1978by,Yang:2023evp}. Because these ingredients are not available for all six bases, absolute couplings are not assigned to the Pauli-filtered spectrum, and values associated with the excluded sextet axial-antidiquark currents are omitted.

We instead define a reduced dynamical amplitude $\gamma_\alpha$ through
\begin{equation}
\overline{|{\cal M}_\alpha|^2}=|\gamma_\alpha|^2
|C_{\alpha}^{\rm color}C_{\alpha}^{\rm spin}C_{\alpha}^{\rm flavor}|^2.
\end{equation}
For an $S$-wave two-body channel, the center-of-mass momentum and width use the standard two-body phase-space normalization~\cite{ParticleDataGroup:2026rpp,Guo:2017jvc}:
\begin{align}
p_\alpha&=\frac{\sqrt{[M_T^2-(m_1+m_2)^2][M_T^2-(m_1-m_2)^2]}}{2M_T},\nonumber\\
\Gamma_\alpha&=\frac{p_\alpha}{8\pi M_T^2}\overline{|{\cal M}_\alpha|^2}
\equiv K_\alpha|\gamma_\alpha|^2.
\label{eq:tq-reduced-width}
\end{align}
When $\gamma_\alpha$ is expressed in GeV, $K_\alpha$ has units of MeV~GeV$^{-2}$. The full ground-state calculation is reported through the threshold excess $Q$, center-of-mass momentum $p$, and reduced coefficient $K_\alpha$, but these channel-by-channel numbers are moved to Supplemental Tables~S7 and S8. The main text therefore summarizes the physically robust fall-apart information in prose rather than introducing another numerical table.

\subsection{Ground-state tetraquark decay spectrum}
\label{subsec:tq_decay_spectrum}
For the antitriplet axial--axial basis $(S_d,S_{\bar d})=(1,1)$, the $0^+$ state contains both pseudoscalar--pseudoscalar ($PP$) and vector--vector ($VV$) recoupling components. Consequently, $D_s\eta^{(\prime)}$ and, when kinematically open, $D_s^*\phi$ probe complementary parts of the same compact color--spin wave function. The corresponding axial--axial $1^+$ state instead couples through $PV$ and $VP$, while its leading $VV$ coefficient vanishes exactly. This $VV=0$ result is an algebraic color--spin selection rule and not a consequence of limited phase space. The mixed-spin antitriplet $(0,1)$ $1^+$ basis has nonzero $PV$, $VP$, and $VV$ components, so it can populate a broader set of daughter-spin combinations, although the physical partial widths remain dependent on the unknown channel couplings.

The surviving sextet bases show the same distinction between recoupling information and dynamical widths. The $(0,0)$ $0^+$ configuration contains $PP$ together with a comparatively enhanced $VV$ projector, while the $(1,0)$ $1^+$ configuration admits $PV$, $VP$, and $VV$. Their larger reduced coefficients should therefore be interpreted as consequences of color--spin recoupling and the available phase space, not as model-independent predictions of broader tetraquark poles; $K_\alpha$ remains only a reduced diagnostic until the corresponding strong couplings are determined.

The antitriplet axial--axial $2^+$ state is especially clean because its leading fall-apart content is $VV$ only. Its physical channel is therefore $D_s^*\phi$, which is open in the SR spectrum but closed in the NR spectrum. This qualitative threshold switch is the sharpest decay-level discriminator between the two kinematic prescriptions. The complete channel-by-channel values of $Q$, $p$, and $K_\alpha$ are retained in Supplemental Tables~S7 and S8 so that the numerical information remains available without obscuring these selection-rule and threshold conclusions in the main text.

The one-daughter cascade construction uses the usual narrow-width product rule~\cite{ParticleDataGroup:2026rpp}:
\begin{align}
\Gamma\big(T\to[H_a\to f_a]H_b\big)&=\Gamma(T\to H_aH_b)\,{\cal B}(H_a\to f_a),\nonumber\\
\Gamma\big(T\to H_a[H_b\to f_b]\big)&=\Gamma(T\to H_aH_b)\,{\cal B}(H_b\to f_b),
\label{eq:tq_cascade_width}
\end{align}
and is tabulated through $K_{\alpha\to f}=K_\alpha{\cal B}(H\to f)$. Only one daughter decays in each row, so the entries are exclusive reconstruction weights rather than additional contributions to the parent width.

\subsection{Representative cascade reconstruction}
\label{subsec:cascade-coefficients}
The complete one-daughter cascade coefficients are moved to Supplemental Tables~S1--S5. In the main text, their role is illustrated by a single chain. If a primary mode $T\to D_s^{*+}X_s$ is open, reconstruction through the dominant radiative daughter decay obeys
\begin{equation}
K_{\rm rec}\big(T\to[D_s^{*+}\to D_s^+\gamma]X_s\big)
=K_{D_s^*X_s}\,{\cal B}(D_s^{*+}\to D_s^+\gamma),
\label{eq:cascade-worked}
\end{equation}
with ${\cal B}(D_s^{*+}\to D_s^+\gamma)=(93.6\pm0.4)\%$~\cite{ParticleDataGroup:2026rpp}. For a produced sample $N_T$, an efficiency $\epsilon$, and a physical parent width $\Gamma_T$, the reconstructed yield scales as $N_{\rm rec}\propto N_T\epsilon K_{\rm rec}|\gamma_\alpha|^2/\Gamma_T$. This relation makes explicit why the cascade coefficients are search weights rather than branching-fraction predictions: the unknown primary strong coupling and total width remain essential. The dominant practical consequence is that open $D_s^{*+}$ channels should preferentially be tagged through $D_s^+\gamma$, while extremely small daughter modes add little experimental reach.

\FloatBarrier
\section{Physical interpretation and robustness of the predictions}
\label{sec:results_discussion}

The calculation connects two complementary parts of the analysis. The conventional $D_s$ sector tests the common model parameters and supplies the daughter masses, form factors, and branching fractions, whereas the Pauli-filtered diquark--antidiquark sector determines the compact $cs\bar{s}\bar{s}$ spectrum and its experimentally relevant threshold and decay patterns. The discussion is organized by robustness rather than by the order in which numbers were generated. First are symmetry-controlled statements: the Pauli-allowed basis, the missing sextet axial-antidiquark family, and exact recoupling zeros. Second are scheme-stable qualitative trends within the present Hamiltonian, especially the antitriplet--sextet hierarchy and broad spin/orbital ordering. Third are quantitatively model-dependent outputs such as absolute masses, threshold excesses, and Regge slopes. Last are exploratory quantities---reduced fall-apart coefficients and candidate mass associations---that additionally depend on unknown strong couplings or omitted mixing. This hierarchy is the basis for deciding which numerical material remains in the main article and which is relegated to the Supplement.

\subsection{The \texorpdfstring{$D_s$}{Ds} meson sector}
\label{subsec:results_ds_meson}

The low-lying spectrum in Table~\ref{mass_meson} provides a useful benchmark because the same charm--strange dynamics enters the tetraquark calculation. Its truncated presentation ($4S$, $3P$, $2D$, $1F$) is intentional even though the calculation extends through $4S$, $4P$, $4D$, and $4F$: the full set is retained for trajectory construction, but only the levels needed to establish calibration quality and spectral ordering are displayed. The SR result for the $1\,{}^{1}S_0$ state, $1967.9$~MeV, differs from the PDG mass by less than $1$~MeV, while the NR value is higher by about $35$~MeV. For the $1\,{}^{3}S_1$ state, the SR and NR deviations are approximately $1$ and $16$~MeV, respectively. The lowest $P$-wave states occupy the expected $2.38$--$2.53$~GeV interval. In the unmixed $LS$ basis the $1\,{}^{3}P_1$ state is reproduced particularly well by SR, whereas the $1\,{}^{1}P_1$ and $1\,{}^{3}P_2$ levels lie below the corresponding experimental masses. This pattern is not unexpected because the physical heavy--light axial states need not coincide with pure ${}^{1}P_1$ and ${}^{3}P_1$ configurations~\cite{Godfrey:1985xj,Ebert:2009ua,Godfrey:2015dva,Ni:2021pce}. The ground-state agreement and conventional orbital ordering indicate that the model is adequate for establishing the mass scale and daughter thresholds, while the residual $P$-wave discrepancies quantify the uncertainty associated with neglected configuration mixing.

The proposed excited assignments show different levels of support. The $D_{s0}(2590)^+$ central value lies between the calculated $2\,{}^{1}S_0$ masses: it is $30.9$~MeV above the SR prediction and only $8.2$~MeV below the NR result. It is therefore compatible with a predominantly radial pseudoscalar interpretation. By contrast, $D_{s1}^{*}(2860)^+$ lies about $130$~MeV above the SR and $102$~MeV above the NR $1\,{}^{3}D_1$ value, so the proposed $D$-wave assignment should be regarded as qualitative until mixing and coupled-channel effects are included~\cite{ParticleDataGroup:2026rpp,LHCb:2020Ds0,LHCb:2014Ds2860,Guo:2017jvc,Moir:2016srx}. The Regge fits reinforce this conclusion at the global level. The orbital slopes are mostly $0.61$--$0.78~\mathrm{GeV}^{-2}$ and the radial slopes $0.42$--$0.49~\mathrm{GeV}^{-2}$, with closely spaced SR and NR values. Thus, the kinetic prescription changes individual excited masses more noticeably than it changes the overall radial and orbital ordering.

The weak-decay calculation is retained only as a validation layer because measured daughter branching fractions enter the reconstruction weights. The representative modes in Table~\ref{tab:ds_decay_combined} show the pattern without reproducing the full numerical inventory: SR and NR are typically within about $10$--$20\%$ for the benchmark $\eta e\nu$ and $\phi e\nu$ channels, while $\eta'e\nu$ is underestimated and the factorized nonleptonic description is visibly less uniform. For example, $\eta\rho^+$ and $\phi\pi^+$ are reasonably close to experiment, whereas $\eta\pi^+$ and $\phi\rho^+$ are overestimated and $\eta'\rho^+$ is strongly underestimated. The complete list, including product branching fractions and upper limits, is therefore kept in Supplemental Table~S6 rather than allowed to dominate the main discussion. This channel dependence is consistent with the known limitations of simple factorization in charm decays~\cite{Bauer:1987BSW,Yu:2023galkin} and confirms that the weak-decay sector is used here as a source of daughter-decay weights, not a precision fit. The experimentally robust input for tetraquark reconstruction is the hierarchy $\mathcal B(D_s^{*+}\to D_s^+\gamma)=(93.6\pm0.4)\%$ versus $\mathcal B(D_s^{*+}\to D_s^+\pi^0)=(5.77\pm0.35)\%$, which strongly favors the radiative feed-down tag~\cite{ParticleDataGroup:2026rpp}.

Overall, the meson analysis does not require exact agreement in every excited or nonleptonic channel to serve its purpose. The accurately reproduced ground states fix the relevant scale, the form-factor calculation provides internally consistent daughter branching fractions, and the measured $D_s^{*+}$ hierarchy identifies the most efficient cascade signatures. These inputs can therefore be used to interpret the tetraquark results while keeping their known systematic limitations explicit.

\subsection{The \texorpdfstring{$cs\bar{s}\bar{s}$}{csss} tetraquark sector}
\label{subsec:results_tetraquark}

The ground-state spectrum displays a pronounced separation between the two color sectors. In the $\bar{\mathbf3}_c\otimes\mathbf3_c$ sector, the axial--axial family gives $0^+$, $1^+$, and $2^+$ states at $2951.1$, $3032.7$, and $3195.8$~MeV in SR and at $2717.9$, $2827.8$, and $3047.4$~MeV in the NR treatment. The mixed-spin $(S_d,S_{\bar d})=(0,1)$ basis produces an additional $1^+$ state at $3089.2$~MeV in SR and $2919.7$~MeV in the NR treatment. The SR spin splittings within the axial--axial multiplet are about $82$~MeV between $0^+$ and $1^+$ and $163$~MeV between $1^+$ and $2^+$; the NR splittings increase to approximately $110$ and $220$~MeV. Hence both schemes preserve the ordering $0^+<1^+<2^+$, but the NR treatment produces a wider multiplet. The allowed sextet $0^+$ and $1^+$ bases cluster near $4.40$~GeV in SR and $4.05$~GeV in NR, roughly $1.2$--$1.4$~GeV above the phenomenologically relevant antitriplet states. This large separation is more important than the small scalar--vector splitting within the sextet pair and makes a low-lying sextet interpretation unlikely in the present Hamiltonian; the broad model dependence of compact tetraquark mass scales motivates comparison with independent constituent and sum-rule calculations~\cite{Ebert:2010af,Zhang:2006ix,Yang:2023evp}.

The radial and orbital spectra add two useful observations. The displayed $3S$, $2P$, and $1D$ subset is a presentation choice rather than a calculation cutoff: the underlying Pauli-filtered spectrum was evaluated through $3S$, $3P$, and $3D$, with the higher omitted towers retained as consistency data. First, the antitriplet radial spacings are not uniform: the first excitation typically lies $0.55$--$0.73$~GeV above the ground state, whereas the second increment is smaller, about $0.36$--$0.41$~GeV. This compression is reflected in the approximately linear, but not perfectly universal, radial Regge fits. Second, the $n=1$ $P$-wave antitriplet states populate roughly $3.46$--$3.68$~GeV in SR and $3.26$--$3.48$~GeV in NR, while the $D$-wave levels form a narrower band near $3.83$--$3.89$ and $3.64$--$3.69$~GeV. The detailed fine structure depends on $S_T$ and $J$, but the broad ordering is stable. The antitriplet radial slopes, $0.255$--$0.317~\mathrm{GeV}^{-2}$, and orbital slopes, $0.308$--$0.467~\mathrm{GeV}^{-2}$, have good linear correlations. The much steeper sextet orbital slopes arise from its compressed heavy-mass sequence and should be treated as a feature of the effective model rather than a universal Regge parameter~\cite{Patel:2025rsf,Patel:2025bmc,Patel:2025bdu}.

Threshold placement converts these mass differences into qualitatively different phenomenology. The antitriplet axial--axial $0^+$ state lies $434.9$~MeV above $D_s\eta$ in SR and $201.7$~MeV above it in the NR treatment. The axial--axial $1^+$ and mixed-spin $1^+$ states are also above their lowest $D_s^*\eta$ thresholds in both schemes, with excesses of $372.6/167.7$ and $429.1/259.6$~MeV. These sizable phase spaces imply that the states need not be narrow merely because they are compact; proximity to threshold, available partial waves, and channel dynamics must be considered together~\cite{Guo:2017jvc}. The $2^+$ basis is exceptional: it lies $64.1$~MeV above the $D_s^*\phi$ threshold in SR but $84.3$~MeV below it in the NR treatment. Observation or non-observation of a near-threshold $D_s^*\phi$ signal would therefore discriminate sharply between the two mass prescriptions. All tabulated $n=1$ $P$- and $D$-wave states are above their assigned lowest spin-compatible thresholds; the nearest $P$-wave case still has $130.1$~MeV of NR excess energy. Their experimental widths will consequently depend on angular-momentum barriers and channel couplings, not on kinematic closure alone.

A direct comparison with other calculations is summarized in Table~\ref{tab:theory-comparison}. The spread is large enough that mass agreement alone is not a structural discriminator. The chiral SU(3) calculation of Zhang \textit{et al.} produces a low mixed scalar eigenvalue near $2.729$~GeV, close to the present NR axial--axial scalar, whereas a recent MIT-bag calculation of the exact $sc\bar s\bar s$ flavor finds its lowest $0^+$ around $3.10$~GeV and several $1^+$ levels between about $3.15$ and $3.38$~GeV~\cite{Zhang:2006ix,Liu:2026compact}. A four-body Gaussian-expansion/complex-scaling study instead reports no bound state below threshold and compact charm resonances primarily in the $3.7$--$3.9$~GeV region~\cite{Zheng:2025csss}. The $2.92\pm0.12$~GeV axial--axial QCD-sum-rule value often used for comparison comes from a related open-flavor current with different light flavor and is therefore an operator analogue, not an exact $cs\bar s\bar s$ prediction~\cite{Yang:2023evp}. Molecular calculations near $2.9$~GeV likewise address different flavor/threshold systems and should be compared through their channel pattern rather than by mass coincidence alone~\cite{Wang:2023eng}.

\begin{table}[!htbp]
\centering
\caption{Selected theoretical mass scales relevant to $cs\bar s\bar s$. ``Exact flavor'' means that the calculation explicitly contains $cs\bar s\bar s$ (or $sc\bar s\bar s$); analogous-current and molecular entries have different light-flavor content and are included only to show model dependence.}
\label{tab:theory-comparison}
\tablefont
\setlength{\tabcolsep}{3.2pt}
\begin{tabular}{@{}L{35mm}L{22mm}L{34mm}L{54mm}@{}}
\toprule
Approach & Flavor relation & Representative mass scale (GeV) & Interpretation \\
\midrule
Present potential model & Exact flavor & $0^+_{AA}:~2.951$ (SR), $2.718$ (NR) & Unmixed diagonal compact basis.\\
Chiral SU(3) quark model~\cite{Zhang:2006ix} & Includes exact flavor with mixing & $0^+:~2.729$ & Mixed $cn\bar n\bar s$--$cs\bar s\bar s$ eigenstate.\\
MIT bag model~\cite{Liu:2026compact} & Exact flavor & $0^+:~3.097,3.441$; $1^+:~3.149$--$3.384$; $2^+:~3.351$ & Color mixing and compactness criterion; high-spin compactness can be disfavored.\\
Four-body GEM+CSM~\cite{Zheng:2025csss} & Exact flavor & resonances mainly $3.7$--$3.9$ & No bound state below the lowest threshold; pole analysis includes four-body dynamics.\\
QCD sum rule~\cite{Yang:2023evp} & Analogous current, different flavor & axial--axial scalar $2.92\pm0.12$ & Useful for current-structure comparison, not an exact-flavor mass prediction.\\
Hadronic molecule~\cite{Wang:2023eng} & Different flavor/threshold system & near $2.9$ & Threshold-organized partner pattern rather than compact color--spin multiplets.\\
\bottomrule
\end{tabular}
\end{table}

No dedicated lattice-QCD pole calculation for the exact $cs\bar s\bar s$ channel was identified in the literature surveyed for this revision. This absence is important: the lattice results cited for conventional $D_s$ states constrain the calibration sector, not the tetraquark pole spectrum. The comparison in Table~\ref{tab:theory-comparison} therefore emphasizes how strongly the predicted mass scale changes when color mixing, four-body dynamics, or threshold degrees of freedom are treated differently.

The physical daughter combinations make the recoupling pattern experimentally testable; analogous use of Fierz/recoupling information to identify preferred fall-apart channels is standard in compact tetraquark decay analyses~\cite{Chen:2020fierz}. Scalar bases can populate $D_s\eta^{(\prime)}$ and $D_s^*\phi$; the axial--axial $1^+$ basis selects $D_s\phi$ and $D_s^*\eta^{(\prime)}$, while the mixed-spin $1^+$ bases also admit $D_s^*\phi$; and the leading $2^+$ channel is $D_s^*\phi$. For the reporting convention $\gamma_\alpha=1$~GeV (used only to expose the scale of $K_\alpha$, not as a physical coupling estimate), the antitriplet sums of reduced coefficients range from $0.079$ to $0.237$~MeV in NR and from $0.257$ to $0.433$~MeV in SR; the SR $2^+$ value is $0.389$~MeV, whereas its NR counterpart vanishes because $D_s^*\phi$ is closed. The sextet sums near $2.1$~MeV chiefly reflect larger phase space and the singlet--singlet color projection, not a model-independent prediction of broader resonances. The Supplemental cascade tables translate each open primary mode into reconstruction tags, usually dominated by $D_s^*\to D_s\gamma$ when a vector charm--strange daughter is present. A recent exact-flavor MIT-bag study reports a different reduced diagnostic, $k|c_i|^2$, based on final-state momentum and color-singlet probability~\cite{Liu:2026compact}. It is conceptually similar to the phase-space/recoupling part of $K_\alpha$, but the operator and normalization conventions differ, so a numerical identification would be misleading. The four-body complex-scaling study instead quotes pole widths for higher resonances~\cite{Zheng:2025csss}; these include dynamics absent from the present reduced-coefficient construction.

\phantomsection\label{subsec:tq-candidate-audit}
$D_{sJ}(3040)^+$ is best regarded as a mass-level benchmark rather than an identified member of the compact spectrum. The PDG lists $I=0$, undetermined $J^P$, $M=3044^{+31}_{-9}$~MeV, and $\Gamma=240\pm60$~MeV; BABAR obtained $M=3044\pm8{}^{+30}_{-5}$~MeV and $\Gamma=239\pm35{}^{+46}_{-42}$~MeV in inclusive $D^*K$ production~\cite{BaBar:2009rro,ParticleDataGroup:2026rpp}. Its central mass lies $11.3$~MeV above the diagonal SR axial--axial $1^+$ basis and $3.4$~MeV below the diagonal NR $2^+$ basis. These numerical coincidences are not robust state assignments because the $1^+$ sector has uncalculated mixing and the two schemes imply different $D_s^*\phi$ threshold status.

The experimental decay information is an equally important constraint. The observed $D^*K$ mode does not belong to the leading flavor-preserving fall-apart topology $cs\bar s\bar s\to(c\bar s)(s\bar s)$ used here; generating $D^*K$ requires additional light-flavor pair creation or other dynamics outside the reduced rearrangement calculation. Moreover, the large measured width cannot be tested against the tabulated $K_\alpha$ values because the channel-dependent $|\gamma_\alpha|^2$ is unknown. We therefore do not claim that the broad width supports a compact tetraquark. A meaningful discriminator would be a measured $J^P$ together with searches for the predicted hidden-strangeness modes $D_s^{*+}\eta$, $D_s^+\phi$, $D_s^{(*)+}\eta'$, and $D_s^{*+}\phi$, preferably as branching-ratio patterns relative to the observed $D^*K$ channel. Until such information exists, a conventional excited $c\bar s$ interpretation remains at least as viable as the compact assignment, given the established quark-model and coupled-channel descriptions of excited charm--strange mesons~\cite{Godfrey:1985xj,Ebert:2009ua,Guo:2017jvc,Moir:2016srx}.

The established structures near $2.9$~GeV are useful analogues but not members of this spectrum. The scalar $T_{\bar c\bar s0}^{*}(2870)^0$ has $M=2874\pm11$~MeV and $\Gamma=68\pm17$~MeV and is observed in $D^-K^+$ and $\bar D^0K_S^0$; its reversed charm--strangeness orientation excludes a direct $cs\bar s\bar s$ assignment~\cite{LHCb:2020bls,LHCb:2020pxc,LHCb:2024xyx,LHCb:2024vfz,ParticleDataGroup:2026rpp}. The vector $T_{\bar c\bar s1}^{*}(2900)^0$, with $M=2904\pm5$~MeV and $\Gamma=106\pm10$~MeV, is additionally far below the predicted negative-parity tower. The neutral and doubly charged $T_{c\bar s0}^{*}(2900)$ states have the appropriate charm orientation but form an isovector $c\bar s q\bar q$ multiplet observed in $D_s^+\pi^-$ and $D_s^+\pi^+$. Their isospin-constrained mass and width, $2908\pm23$ and $136\pm28$~MeV, support a similar open-charm scalar mass scale but not the same flavor assignment~\cite{LHCb:2023Tcbs2900,ParticleDataGroup:2026rpp}.

A compact assignment therefore requires simultaneous agreement in conserved quantum numbers, mass, and the correlated hidden-strangeness decay pattern. Among listed resonances, only $D_{sJ}(3040)^+$ presently satisfies the first two criteria; the channels identified above test the third.

\section{Conclusion}
The most model-independent result of this work is the Pauli filtering of the internally $S$-wave $\bar s\bar s$ pair: it fixes the allowed antidiquark color--spin combinations, removes the sextet axial-antidiquark family, and produces exact recoupling selection rules for the leading fall-apart topology. The numerical spectrum is less robust. The antitriplet states occupy the phenomenologically accessible region in both SR and NR treatments, while the surviving sextet bases are much heavier, but the absolute masses, threshold placement, and Regge slopes show substantial scheme and model dependence. In particular, the close diagonal $1^+$ levels make off-diagonal mixing a priority for any state-level interpretation.

The decay calculation should be read as a hierarchy of kinematic and algebraic diagnostics rather than a prediction of absolute widths. The coefficients $K_\alpha$ specify phase space and color--spin--flavor recoupling for a stated normalization, while the physical channel-dependent coupling remains to be calculated. The especially different SR and NR placement of the $2^+$ basis relative to $D_s^*\phi$ is a useful discriminant. For the same reason, exhaustive decay coefficients, the full $D_s$ weak-decay inventory, and calibration Regge plots are placed in Supplemental Material, while the main article retains only the quantities needed to establish the physical argument. Likewise, although the tetraquark spectrum was calculated through $3S$, $3P$, and $3D$, only $3S$, $2P$, and $1D$ are displayed; the meson calculation extends through $4S$, $4P$, $4D$, and $4F$, while its main table is limited to $4S$, $3P$, $2D$, and $1F$. These truncations reduce numerical overload without changing the calculations or the stated conclusions.

$D_{sJ}(3040)^+$ is consequently retained only as a mass-level benchmark. Its broad width, unknown $J^P$, observed $D^*K$ decay, and the unresolved compact-basis mixing prevent a robust tetraquark assignment. The next decisive calculations are the $0^+$/$1^+$ mixing matrices and current-specific strong couplings (or a coupled-channel pole analysis); experimentally, $J^P$ determination and correlated $D_s^{(*)}\eta^{(\prime)}$ and $D_s^{(*)}\phi$ searches provide the clearest tests.

\sloppy\setlength{\emergencystretch}{2em}\FloatBarrier
\begin{spacing}{1.15}
\setlength{\hfuzz}{2pt}

\end{spacing}

\end{document}